\documentclass[12pt, preprint]{aastex}

\shorttitle{Effective Temperatures of Red Supergiants}
\shortauthors{Levesque et al.}

\begin{document}

\title{The Effective Temperatures and Physical Properties of Magellanic Cloud Red Supergiants:
The Effects of Metallicity}

\author{Emily M. Levesque\altaffilmark{1,2}, Philip Massey\altaffilmark{2}}

\affil{Lowell Observatory, 1400 W. Mars Hill Road, Flagstaff, AZ 86001}
\email{emsque@mit.edu, Phil.Massey@lowell.edu}

\author{K. A. G. Olsen\altaffilmark{2}}
\affil{Cerro Tololo Inter-American Observatory, National Optical Astronomy Observatory, Casilla 603, La Serena, Chile}
\email{kolsen@noao.edu}

\author{Bertrand Plez}
\affil{GRAAL CNRS UMR5024, Universit\'{e} de Montpellier II, 34095 Montpellier Cedex 05, France}
\email{Bertrand.Plez@graal.univ-montp2.fr}

\author{Georges Meynet and Andre Maeder}
\affil{Geneva Observatory, 1290 Sauverny, Switzerland}
\email{georges.meynet@obs.unige.ch,andre.maeder@obs.unige.ch}

\altaffiltext{1} {Current address:
Massachusetts Institute of Technology, 77 Massachusetts Avenue, Cambridge, MA 02139.}
\altaffiltext{2} {Visiting Astronomer, Cerro Tololo Inter-American Observatory (CTIO), National Optical 
Astronomy Observatory (NOAO), which is operated by the Association
 of Universities for Research in Astronomy (AURA), Inc., under cooperative agreement
with the National Science Foundation (NSF).}

\clearpage

\begin{abstract}

We present moderate-resolution optical spectrophotometry of 36 red
supergiants (RSGs) in the LMC and 39 RSGs in the SMC. Using the
MARCS stellar atmosphere models to fit this spectrophotometry, we
determine  the reddenings, effective temperatures and other physical
properties, such as bolometric luminosity and effective stellar
radii, and compare these to stellar evolutionary models.  As a
self-consistency check, we also compare the broad-band colors
$(V-K)_0$ and $(V-R)_0$ with the models.   The $(V-R)_0$ results
are in good agreement with those from fitting the optical
spectrophotometry, but the $(V-K)_0$ results show metallicity-dependent
systematic differences, amounting to 3-4\% in effective temperature,
and 0.2~mag in bolometric luminosity, at the metallicity of the
SMC; we conclude that this is likely due to the limitations of
static 1D models, as spectra of RSGs in the optical and IR may
reflect different atmospheric conditions due to the large surface
granulation present in these stars.  We adopt the scales indicated
by the optical spectrophotometry and $(V-R)_0$ colors, but accept
that there is still some uncertainty in the absolute temperature
scales.  We find that the effective temperature scales for the LMC
and SMC  K-type supergiants agree with each other and with that of
the Milky Way, while for M-type supergiants the scales are cooler
than the Galactic scale by 50~K and 150~K, respectively.  This is
in the sense that one would expect: since the spectral classification
of RSGs is based on the line strengths of TiO, stars with lower
abundances of these elements have to be cooler in order to have the
same strength.  However, this effect is not sufficient to explain
the shift in average RSG spectral type between the three galaxies.
Instead, it is the effect that metallicity has on the coolest extent
of the evolution of a star that is primarily responsible.  Our new
results bring the RSGs into much better agreement with stellar
evolutionary theory, although the SMC RSGs show a considerably
larger spread in effective temperatures at a given luminosity than
do the LMC stars.  This is expected due to the larger effects of
rotational mixing in lower-metallicity stars, as higher helium
abundance at the surface would lead to higher effective temperatures
in the RSG phase.  We also find that the distribution of reddening
of RSGs in the Clouds is skewed significantly towards higher values,
consistent with our recent finding that Galactic RSGs show extra
extinction due to circumstellar dust.

\end{abstract}


\section{Introduction}
\label{Sec-intro}

Until recently, the location of  Galactic red supergiants (RSGs)
 in the H-R diagram was poorly matched by 
stellar evolutionary tracks (Massey 2003), with evolutionary
theory failing to produce stars as cool and luminous as those ``observed."  Many
possible explanations might contribute to this discrepancy: there is poor knowledge of
RSG molecular opacities, the near-sonic velocities of the convective layers 
invalidate simplifications of mixing length theory, and the highly extended
atmospheres of these stars differ from the plane-parallel geometry assumption
adopted by evolutionary models. In truth, the disagreement between theory
and observation lay not in deficiencies of theory, but in an incorrect
placement of RSGs in the H-R diagram. Levesque et al.\  (2005, hereafter Paper~I) used the
new generation of MARCS atmosphere models (Gustafsson et al.\ 1975, 
Plez et al.\ 1992,  Plez 2003, Gustafsson et al.\ 2003)  to fit moderate-resolution optical
spectrophotometry of 74 Galactic RSGs.  The newly derived physical parameters were
in excellent agreement with the Geneva evolutionary tracks for solar metallicity 
(Meynet \& Maeder 2003). 

A similar problem is known to exist for RSGs in the Magellanic Clouds (MCs), as shown
in Fig.~\ref{fig:oldhrd}, where the data are from Massey \& Olsen (2003), and is based upon
the best available calibration at the time.  One can see that the evolutionary tracks do not
extend to cool enough temperatures to reproduce the ``observed" (assumed) location in 
the H-R diagram.  In addition, there is a long-standing mystery as to
why the distribution of RSG spectral subtypes in the MCs is skewed towards earlier types
in the MCs (Elias et al.\ 1985), with the average RSG being K5-K7  in the SMC, M1 in
the LMC, and M2 in the Milky Way (Massey \& Olsen 2003).  The spectral subtype of
late K- and M-type stars is largely determined on the basis of the 
strengths of the TiO bands, which are highly sensitive to temperature, 
but their strengths will obviously
also depend upon the chemical abundances.
Massey \& Olsen (2003) proposed that the change 
in the distribution of spectral types was due to  
the lower abundances found in the Clouds ($Z/Z_\odot=0.2$ for the SMC, and
$Z/Z_\odot$=0.5 for the LMC; see  Westerlund 1997 and discussion in Massey et al.\ 2004);
 i.e., that a 3800~K star would simply appear
to be of earlier spectral type in the SMC because of the lower abundance of TiO. 
Alternatively, as suggested by Elias et al.\ (1985),
it is possible that stars evolve to cooler temperatures at higher metallicities than at lower, since the
``Hayashi limit" (the maximum radius as a function of mass) decreases with
metallicity (Hayashi \& H\={o}shi 1961; see also Sugimoto \& Nomoto 1974).

Understanding the physical properties of red supergiants  at the low metallicities that
characterize the Magellanic Clouds is of particular importance.  Such data
can be combined with that of other galaxies in the Local Group
for a sensitive test of stellar
evolutionary models as a function of metallicity.  Observationally, the relative number
of RSGs and Wolf-Rayet stars appear to change by a factor of $>$100 over 0.8~dex
in metallicity (Massey 2002, 2003), in accordance with the suggestion first made by
Maeder et al.\ (1980).  In addition, the number ratio of RSGs to blue supergiants
will be much higher at low metallicities (van den Bergh 1973), although there are
challenges in determining this ratio quantitatively (Massey 2002).
In the Milky Way,
RSGs contribute only a few percent to the dust content of the interstellar medium, but in starburst
galaxies, or galaxies at large look-back times, we expect that RSGs will play a major role
(Massey et al.\ 2005a), as such galaxies are usually metal-poor.
Finally, the MCs represent a relatively ``clean" environment, with
minimal and uniform reddenings, which avoid some of the difficulties inherent in studying
RSGs in the Milky Way (Paper I).

Oestreicher \& Schmidt-Kaler (1998, 1999) used an earlier version of the MARCS
models (Bessell et al.\ 1998) with their own spectrophotometry and CCD photometry
to analyze a large sample of LMC stars.  They conclude that the MARCS models did
the best of the then-available models at fitting the data, and derived 
physical properties using these fits.  Our work here was partially inspired by this
work, and offers the following improvements.  First, the MARCS models used 
here (and in Paper I) have been substantially revised.  These now include
sphericity, with an order of magnitude increase in the number of opacity
sampling points, incorporate improved atomic opacities (both line and
continuum), and also include improved molecular opacities (e.g., CN, CH).
Second, there is now improved broad-band photometry available for {\it both}
the LMC and SMC thanks to 2MASS and the recent CCD survey of Massey (2002).
Also, our sample consists of stars whose radial velocities have been shown to be consistent
with membership in the Magellanic Clouds by Massey \& Olsen (2003).  This 
list
extends to stars which have cooler effective temperatures than do any of the Oestreicher \&
Schmidt-Kaler (1998, 1999) LMC stars.  In addition, our sample includes SMC
RSGs, providing important tests at even lower metallicities.  Since we use the same
techniques, quality of data, and modeling for RSGs in all three galaxies, differences
in the derived physical properties are likely to reflect real differences, and not just
methodology.

Here we present moderate-resolution spectrophotometry of 36 LMC RSGs and
39 SMC RSGs (\S~\ref{Sec-obs}). From these data we determine spectral
types (\S~\ref{Sec-types}), and derive  physical
parameters of RSGs in each of the Clouds (\S~\ref{Sec-analysis}).  As
a test of the consistency of the MARCS models, we  compare
these to what we would derive purely on the basis of $(V-K)_0$ 
and $(V-R)_0$ colors (\S~\ref{Sec-phot}).
In \S~\ref{Sec-results} we discuss our results:
we compare the extinction found for our
stars to that of OB stars in the Clouds (\S~\ref{Sec-reddening}), compare the newly derived
physical properties to those predicted by stellar evolution theory (\S~\ref{Sec-evol}), and compare the
effective temperature scales for the Magellanic Cloud RSGs to that of the Milky Way (\S~\ref{Sec-met}).
In \S~\ref{Sec-sum} we summarize our results, and lay out the directions
for our future work.

\section{Observations and Reductions}
\label{Sec-obs}

\subsection{Target Selection}

Our stars are listed in Table~\ref{tab:stars}.  The sample is drawn from
Massey \& Olsen (2003), who relied upon radial velocities to distinguish
foreground dwarfs from MC RSGs.  
We expect that contamination by red giants
in the halo
(which would occupy a similar color and magnitude range) 
will be low, only a few percent,
but such stars would be hard to distinguish from MC supergiants,
as most of the radial velocity of the Clouds is simply due to the reflex
motion of the sun.  The optical photometry and positions come from Massey
(2002), while the $K_S$ values are from the Two Micron All Sky Survey (2MASS)
database.  The preliminary spectral types assigned by Massey \& Olsen (2003)
were used to ensure that a good range of spectral subtypes were observed
in each Cloud. Some additional stars were observed in each Cloud, but lacked
complete spectral coverage, and these stars will be discussed separately once complete
data are acquired.  In addition, we observed a few Galactic RSGs from Paper~I to
use in classifying the Magellanic Cloud stars, and to act as a check on our fluxes;
this later turned out to be valuable, as detailed below.

\subsection{Observations}

We obtained spectroscopic data with the R-C Spectrograph on the CTIO Blanco
4-meter telescope  during six nights (UT 2004 November 23-25, 1-2 Dec, 4 Dec),
using the Blue Air Schmidt camera and Loral 3K CCD. We used
a slit width of 375 $\mu$m (2.5"), which projected to roughly 3.8 pixels on
the detector.  In the blue we used a 632 l mm$^{-1}$ grating (``KPGL1")
blazed at 4200\AA\  for coverage from  3550\AA\ - 6420\AA\  in first order at
1.01 \AA\ pixel$^{-1}$, with a 
GG-345 blocking filter to block any second-order light.  In the red
we used a 632 l mm$^{-1}$ grating (''KPGLF") blazed at 8400\AA\ for coverage
6130\AA\ - 9100\AA\  in first order at  1.04 \AA\  pixel$^{-1}$, 
with a GG-495 blocking filter to block any second-order light.
The spectral resolution was 3.8 \AA\ for both setups.  Observations in the blue
were obtained on four nights (23-25 Nov 2004, 2 Dec 2004), and observations in the red on 
two nights (1, 4 Dec 2004).
The chip was binned by 2 pixels in the spatial direction, resulting in a scale
of  1.0 arsec pixel$^{-1}$.    
All observations were made with the slit oriented near the parallactic angle.
Conditions ranged from moderate cirrus to clear during the run, with seeing of 1.2-2.0 arcsec.  Typical exposure times were 300 s in the blue, and 200 s in the red.

Bias frames were obtained each evening after the dewar was
filled.
Flat field calibration was obtained by taking projector flats at the
beginning of each night; dome flats yielded similar solutions.  Wavelength calibration
was obtained by taking exposures of a He-Ne-Ar calibration source throughout the night.
Observations of spectrophotometric standards were made throughout the night,
and we also included in our program several Galactic spectral standards, in common
with our Galactic program (Paper I).

\subsection{Reductions}

We reduced the data using Image Reduction and Analysis Facility 
(IRAF)\footnote{IRAF is distributed by NOAO,  which is
operated by AURA, Inc.,
under cooperative agreement with the NSF.}.   Each frame was corrected
for the bias level by a value determined from the overscan region, and then corrected
for the (negligible) two-dimensional bias structure determined from average bias exposures
for each night. A low-order function was fit in the dispersion direction to normalize the
average of the flat-field exposures, and the normalized flat was then divided into each frame on
a nightly basis.    The spectra were extracted using an optimal extraction algorithm, and then
wavelength corrected.

The spectrophotometric standards were used to construct sensitivity curves for each night.
A grey-shift of each standard was allowed, and typically resulted in an RMS of 0.01-0.02~mag
for the six or seven standard observations made each night.  The standards bracketed the
range of airmasses for which the program objects were observed, and standard values were
assumed for the extinction.  

As noted above, we observed a few Galactic RSGs from Paper~I to act as spectral standards, and
to serve as a consistency check on our data.  
Given the good agreement of the spectrophotometric
standard star observations, we were quite surprised to find
that several of the Galactic RSGs
differed quite significantly in the near ultra-violet (NUV) fluxes, 
particularly below 3800\AA, from what we had found in Paper~I.
The spectrophotometric
standards  agreed very well in this region, 
and yet the same disagreement was seen when comparing the new data to those obtained
on either the Kitt Peak 2.1-m or CTIO 1.5-m telescopes (Paper I).  We finally determined
that the problem was inherent to the data, and not the reduction techniques.
The new data all had extra flux in the NUV.  We eventually noticed a strong correlation with color:
the reddest red supergiants had the largest discrepancy.  We also found that there was
unexpected structure to the NUV flux, and in particular that there was a feature
at 3810\AA\ which looked remarkably like the telluric A-band at 7620\AA,  i.e., at exactly
twice the wavelength of the NUV feature.  The conclusion was obvious: somehow
the flux at a given
wavelength was being affected by the flux at twice the wavelength\footnote{We are indebted to a colleague who, upon hearing of this problem, dubbed it ``the old red-leak gotcha".}.
Of course, since we had been observing with a first-order grating in the blue, this is not the sort
of typical order-separation problem that come with the lack or misuse of blocking filters.
By subtracting our spectrophotometry obtained in Paper I from the current data, and comparing
that to the counts in the red, we established that there was a few percent ghost of 2$\lambda$
light contaminating our observations.  The problem does not show up with stars of normal
colors (such as that of the spectrophotometric standards), but becomes significant for the
extremely red stars we observed.   Tests by K.A.G.O. and P.M. in March 2005
using the comparison arcs and various blocking filters established beyond any doubt that
this 632 l mm$^{-1}$ grating also acts as a 316 l mm$^{-1}$ grating, albeit at a low level.
Another replica of this grating has been in use for many years with the Kitt Peak 4-m RC spectrograph
(``KPC-007").  After our discovery, Di Harmer kindly conducted a similar test with it, and found that
it suffers from the same problem.

We determined an empirical correction factor, which amounted to several percent of the 2$\lambda$
count-rate, and applied that to all of our data.  The spectrophotometric standards yielded the identical
solutions.  The correction is significant (greater than a few percent) only below 4000\AA.  So as to
not compromise the results of the present study, we restrict ourselves only to data long-wards of 4100\AA.

\section{Analysis}
\label{Sec-analysis}

\subsection{Spectral Types}
\label{Sec-types}

Since RSGs can vary in spectral type, and since our current spectra have higher signal-to-noise
than those of Massey \& Olsen (2003), we chose to reclassify all of the Magellanic Cloud stars
in our sample.  This reclassification was based primarily on the TiO band depths, which
are visible even for the early and mid Ks (specifically the $\lambda 5167$ and $\lambda 6158$ TiO
lines; see also Jaschek \& Jaschek 1987).  We strove for consistency between the classifications
used here and in Paper I. The revised spectral types are compared to the older ones in
Table~\ref{tab:stars}.

\subsection{Modeling the Spectrophotometry}
\label{Sec-teff}

Our fitting of the spectrophotometry determines three properties of these stars:
effective temperature $T_{\rm eff}$, the visual extinction $A_V$, and (indirectly)
the surface gravity $g$ [cgs].
To accomplish this, we compared our observed spectral energy distribution
to a series of MARCS stellar atmosphere models.  The models were
computed for a metallicity $Z/Z_\odot=0.2$ (SMC),
and  $Z/Z_\odot=0.5$ (LMC). Of course, the assumption that
the abundances of all elements scale with a single ``metallicity" value is a simplification,
but may be a good approximation: see, for example Pritzl et al.\ (2005), who find that solar-like
ratios of alpha-products to Fe are common even in very metal-poor systems.
In any event, this serves as
a useful starting point for determining the effect that chemical abundances have
on the physical parameters of these stars. The models ranged from 
3000~K to 4500~K in increments of 100~K and with $\log g$ values from -1 to +1, in increments of 0.5~dex.
We interpolated the models for intermediate temperatures at 25 K increments.
When making the fits, we reddened
the models using a Cardelli et al.\ (1989) reddening law with $R_V=3.1$.  Although a high value
of the {\it effective} ratio of selective-to-total extinction is needed to correct broad-band photometry of
such red stars, the same $R_V$ that works for early-type stars will work for RSGs when dealing
with optical spectrophotometry; see Massey et al.\ (2005a). 

None of the spectral features have an obvious surface gravity dependence, and so we used the same
procedure as in Paper~I to arrive at a final set of values.
We began the fitting procedure with the
$\log g=0.0$ models, and determined the reddening and effective temperatures that gave the
best fit (by eye) both to the spectral features and to the continuum.  These fits were unique and
well-determined, with a precision of 50~K for the M stars and the mid-to-late K stars.  For the earlier K stars, our fits were based primarily
on TiO $\lambda 5167$
and the G-band; we estimate that the effective temperatures of these stars have been obtained to a precision of 100~K.  The extinction values $A_V$ are determined to
0.15~mag.    We next checked to see if the initial $\log g=0.0$ value was appropriate to that expected
for the star: the bolometric corrections (from the models)
 were used with the reddening and photometry (Table~\ref{tab:stars}) to compute the bolometric luminosity, assuming true distance moduli of 18.9 (SMC) or 18.5 (LMC).  The bolometric luminosity
 and effective temperature define an effective radius, which is used with an estimate of the 
 mass (from a simple mass-luminosity relation determined from the Geneva evolutionary
 models, and which remains valid at these lower metallicities) to determine the physical
 $\log g$.  If these $\log g$ values were closer to +0.5 or -0.5 than to our initial estimate of
 0, then the star was refit with a model with a more appropriate value for the surface gravity.
 The calculation was then repeated, although the results converged quickly.
 In practice a difference in the $\log g$ value had no effect on the effective temperature,
 but slightly changed the extinction estimate.  See Paper~I for more details.
We show four sample fits in Fig.~\ref{fig:samples}.  The complete set is available in the on-line
edition,
and we are also making our spectra and the models available through the Centre de Donnees
Astronomiques de Strasbourg's VizieR server\footnote{The opacity-sampled synthetic
spectra need to be smoothed in order to compare them to data.  For the broad
molecular bands the exact smoothing is unimportant, but for comparing atomic lines
the degree of smoothing matters.  For example, the slight differences  between the observed spectra and fitted models visible near 5200\AA\ in Fig.~\ref{fig:samples} 
is likely due to the presence of the Mg I triplet (only partially resolved at our spectral
resolution) superimposed on the TiO $\lambda 5167$.  Decreasing the smoothing
removes most of this slight discrepancy, but introduces other problems due to the
finite sampling in producing the synthetic spectra.  Since this region was not used in fitting the models to the data, the disagreement in this region is cosmetic, and not
indicative of a problem.  We are indebted to the anonymous referee for raising
this issue.}.

 We give our final values in Table~\ref{tab:results}.   
 In determining the bolometric luminosity we used the models to compute the bolometric correction
 at $V$ as a function of effective temperature for each galaxy.

 Milky Way:
 $${\rm BC}_V=-298.954 + 217.532 (T_{\rm eff}/1000~K) -53.1400 (T_{\rm eff}/1000~K)^2 +4.34602 (T_{\rm eff}/1000~K)^3$$

 LMC:
 $${\rm BC}_V=-121.364 + 78.41064 (T_{\rm eff}/1000~K) -16.8979  (T_{\rm eff}/1000~K)^2 +1.20674  (T_{\rm eff}/1000~K)^3$$

 SMC:
 $${\rm BC}_V=-120.102 + 82.3070 (T_{\rm eff}/1000~K)  -19.0865  (T_{\rm eff}/1000~K)^2 +1.48927  (T_{\rm eff}/1000~K)^3$$

\subsection{Analysis of Broad-Band Photometry}

\label{Sec-phot}

Although we expect that the TiO molecular band strengths will be directly affected by the
abundances, and hence that the calibration of effective temperature with spectral
type will depend upon metallicity, it is less clear {\it a priori} 
how the calibration of effective temperature with broad-band photometry will depend upon
the metallicity.  Josselin et al.\ (2000) have emphasized the usefulness of $(V-K)_0$ as a 
temperature indicator for Galactic RSGs, and in Paper~I we showed that reasonable agreement
exists between the physical parameters derived from fitting the spectrophotometry with those
derived from $(V-K)_0$.    In the optical, $(V-R)_0$ is known to be a good temperature indicator,
while $(B-V)_0$ is instead dominated by surface gravity effects (Massey 1998) due to line-blanketing
by weak metal lines.

\subsubsection{Temperatures and Luminosities from $(V-K)_0$}

\label{Sec-vmk}

In Fig.~\ref{fig:vmkfit}(a) we show the fits derived from the synthetic $(V-K)_0$ colors, derived from
our models using the procedure and assumptions of Bessell et al.\ (1998).  As in Paper~I,
the $V$ bandpass comes from Bessell (1990), while the K bandpass comes from Bessell \& Brett (1988).
Solar metallicity is shown in black, with the LMC metallicity indicated by red, and the SMC by green.
The dispersion at each temperature is due to the range of $\log g$.  The three curves show the
fit to all of the data for each galaxy ($T_{\rm eff}$=3200-4300~K, $\log g=-1,-0.5,$0,+0.5, and +1.0,
when available).  For simplicity we have indicated the $\log g=0.0$ models, which are the
most typical of our sample, with filled circles.   The dispersion with $\log g$ clearly increases
with decreasing temperature, but is small for $(V-K)_0\le4.5$ or $T_{\rm eff}\ge$ 3600~K.  In this
regime there is also little difference between the calibrations at different metallicities, with
about a 30~K change (SMC minus Milky Way).  The coolest stars in our Magellanic Cloud sample
have $T_{\rm eff}\sim 3475$, corresponding to $(V-K)_0\sim 5.0$, for which the differences are  -50~K
(SMC minus Milky Way). The formal fits are given here:

Milky Way:
$$T_{\rm eff}=7741.9 -1831.83 (V-K)_0+263.135 (V-K)_0^2-13.1943 (V-K)_0^3$$

LMC:
$$T_{\rm eff}=7621.1-1737.74 (V-K)_0+241.762 (V-K)_0^2-11.8433 (V-K)_0^3$$

SMC:
$$T_{\rm eff}=7167.5-1374.20 (V-K)_0+157.000 (V-K)_0^2-6.0481 (V-K)_0^3$$

\noindent
The RMS of these fits are 11 K (Galactic), 17 K (LMC), and 36 K (SMC), due primarily to
the spread of $\log g$.  However, in practice  none of the stars in our LMC and SMC
samples have extreme $\log g$ values, and as Fig.~\ref{fig:vmkfit}(a) shows, there is good
agreement between our fits and the (usual) $\log g=0.0$ case indicated by the filled circles.  

How does the sensitivity of $T_{\rm eff}$ to color compare to the sensitivity of spectral fitting?
As discussed above, we felt that the typical error for our mid-K to M stars was about 50~K.
A reasonable error for $(V-K)_0$ might be 0.10~mag, given the need to correct for reddening,
and we note that at 3800~K this change in color would correspond to about 35~K, 
smaller but comparable to the uncertainty in our spectral fitting.

Few of our Magellanic Cloud sample have ``standard" K-band photometry, but all
have been observed by the Two Micron All Sky Survey (2MASS) in their  $K_S$ system 
(i.e., the $V-K_S$ values in Table~1).
Paper I cautioned that the two are not quite equivalent,  and we have compared
the ``standard CIT" K-band photometry of Elias et al.\  (1985) with the $K_S$ photometry
found in the 2MASS catalog.  Indeed, there is a small but non-negligible offset, with
$K_{\rm CIT} - K_S=+0.04$.  This value comes from 41 SMC stars, and has a small scatter (RMS of 0.06~mag); it is similar to the conversion found by Carpenter (2001), who find
$K_{\rm CIT} - K_S=+0.024$. Bessell \& Brett (1988)
find that the transformation between their adopted K bandpass (upon which our
synthetic colors are based), and that of the CIT system,
amounts to a constant offset $K=K_{\rm CIT}+0.02$.   We expect therefore
that $K=K_S+0.06$ for our data-set. This is in accordance with the transformation
given by Carpenter (2001), who finds that $K=K_S+0.044$ for 2MASS data.
We have corrected the
$V-K_S$ values appropriately before applying the above conversion to $T_{\rm eff}$.
This correction is small, amounting to 20~K at $(V-K)_0=4.0$.

In Paper~I we assumed that the extinction correction at $K$ was $0.11 A_V$ following
Schlegel et al.\ (1998).   Massey et al.\ (2005a) have (re)emphasized the need to carefully
consider the spectral energy distribution of the source when correcting broad-band photometry
of RSGs (see McCall 2004 for discussion of the general case).  We have re-examined the
issue here for $K_S$ using the MARCS models and a Cardelli et al.\ (1989) reddening
law, and find that over the range of temperatures relevant here we derive a numerical
value in excellent agreement, $A_{K_S}=0.12 A_V$.   Thus we expect $(V-K_S)_0=(V-K_S)-0.88A_V$,
where the $A_V$ values come from Table~\ref{tab:results}. 

In Paper~I we found that the bolometric correction at K for Galactic stars was linear with
respect to effective temperature, with a small dispersion:

Milky Way:
$$BC_K=5.574 - 0.7589(T_{\rm eff}/1000~K).$$

\noindent
Here we find similar results for the LMC and SMC metallicities:

LMC:
$$BC_K=5.502 - 0.7392(T_{\rm eff}/1000~K)$$

SMC:
$$BC_K=5.369-0.7029(T_{\rm eff}/1000~K)$$

\noindent
In all cases the data have been fit over the range 3200-4300~K and for $\log g=-1$ to +1.
The dispersion is 0.01 for the Galactic and LMC metallicity models, and 0.02 mag for the SMC,
where again the dispersion is mostly due to the effects of extreme surface gravities (+1, -1) at
low effective temperatures.  We show the fits in Fig.~\ref{fig:vmkfit}(b).

We list the derived effective temperatures in Table~\ref{tab:bb}, and
compare these to those derived from fitting the spectrophotometry in Fig.~\ref{fig:diff}(a) . 
There is clearly a systematic difference, with
the $(V-K)_0$ relation predicting a higher effective temperature than
the spectral fitting.   We saw a similar
effect with the Galactic stars in Paper~I (their
Fig.~5a), although the scatter was significant, due presumably
to the large (and in some cases, uncertain) reddening.  Here the stars are lightly reddened, and
the offset more obvious.   Nevertheless, the median difference for the Galactic sample 
(in the sense of  spectral optical minus IR)  is the same as for the LMC data
here, -105 K.  The median difference for the SMC stars is -170 K.

This temperature difference will of course translate into a difference in the bolometric luminosities
for these stars, as the bolometric correction is a steep function of effective temperature.  We list
the derived luminosities in Table~\ref{tab:bb}, and show the comparison with those derived
from fitting the spectrophotometry in  Fig.~\ref{fig:diff}(b).  The differences amount to about -0.20~mag at the metallicity of the 
SMC.

This suggests that there is an inconsistency in the IR fluxes predicted by the models for
a given TiO band depth. The difference amount to about
0.5~mag at $K$ at low metallicities ($Z/Z_\odot=0.2$), with better agreement at higher metallicities.  What could be the cause? 
Josselin \&  Plez (2005) showed that low excitation IR CO lines in the K band
computed using MARCS models are too faint compared to observations
of Betelgeuse. Also, Alvarez et al.\ (2000) , point out that their CO index is larger in RSG than in other
late-type giants. However, this CO absorption not accounted for by the models
cannot be responsible for the difference of 40\% 
in the K flux we find here,
and seems to exclude an unknown opacity source in the K band.
Rather, the explanation might lie in the shortcomings of 1D static models.
It was shown recently (Ryde et al.\ 2005)
that MARCS models at
the canonical $T_{\rm eff}$ =3600~K for Betelgeuse
cannot reproduce the IR H$_2$O lines around 12$\mu$m, while spectra generated
with a $T_{\rm eff}$=3250~K do reproduce the observations. In the optical a temperature
of 3600~K is more appropriate. Radiative-hydrodynamical 3D models
of RSG do show a pattern of large warm and cool patches on the surface (Freytag et al. 2002)
that may explain this wavelength dependent $T_{\rm eff}$.
We can readily provide a rough estimate, at the SMC metallicity, of the impact on
$(V-K)_0$ of an optical spectrum characterized by a $T_{\rm eff}$=3600~K  
(($V-K)_0=4.49$)
and an IR spectrum characteristic of a lower 
$T_{\rm eff}$=3200~K ($(V-K)_0=6.56$).
Using two MARCs models at these $T_{\rm eff}$
and taking their $V$ and $K$ magnitudes, we find a composite $(V-K)_0= 4.24$, which
corresponds to a MARCS model at 3703~K. Thus, using the $(V-K)_0$ would lead to a
$T_{\rm eff}$=3700~K, while the fitting of TiO bands would lead to $T_{\rm eff}$=3600~K. This explains
a large part of the effect we observe, and would also impact the bolometric correction,
but  more detailed calculations
are necessary with 3D models, or at least with patches of 1D models.

\subsubsection{Temperatures and Luminosities from $(V-R)_0$}

\label{Sec-vmr}

Of the optical colors, $(V-R)_0$ has the greatest potential for being used for a temperature
indicator; see discussion in Massey (1998).   By comparing the temperatures arrived at by $(V-R)_0$ we may also obtain useful clues as to the source of the discrepancy we found when comparing the
effective temperatures derived from $(V-K)_0$ (\S~\ref{Sec-vmr}) with those found from fitting
the optical spectrophotometry (\S~\ref{Sec-teff}).

Using the 
Bessell (1990) approximation for the (Johnson) $V$ and  (Cousins) $R$ filters, we have computed the expected colors for each of the MARCS models.
The following fits have been made discarding the stars of extreme surface gravities
($\log g$=+1, -1):

Galactic:
$$T_{\rm eff}=8304.4 -9158.6 (V-R)_0 + 5675.2 (V-R)_0^2 - 1194.90 (V-R)_0^3$$

LMC:
$$T_{\rm eff}=7798.3 - 7824.4 (V-R)_0 + 4554.8 (V-R)_0^2 -905.21 (V-R)_0^3$$

SMC:
$$T_{\rm eff}=7179.4 - 6030.8 (V-R)_0 + 3028.2 (V-R)_0^2 -525.98 (V-R)_0^3$$

\noindent
where the dispersions are  35 K, 40~K, and   68 K, respectively.  We emphasize that these
fits only apply to the range $T_{\rm eff}>3200$K, which corresponds roughly to $(V-R)_0<1.8$
for the Milky Way and LMC, and $(V-R)_0 < 1.5$ for the SMC.   The fits are shown
in Fig.~\ref{fig:vmrfit}.  The sensitivity is less than that with $(V-K)_0$, as might be expected
given the smaller baseline.  An error of 0.05 mag in $(V-R)_0$ is not unreasonable, and
would amount to an error of 90~K.  The derived relationship for the LMC can be
compared to
that found by Oestricher \& Schmidt-Kaler (1999) from the older models; the scatter
from their fitting is twice as great. The temperatures agree for the warmer
RSGs (50~K difference at $(V-K)_0$=1.0) but disagree considerably for the coolest
RSGs (200 K difference at $(V-K)_0$=1.5).

According to Schlegel et al.\ (1998), we can expect that $A_R=0.81 A_V$ for the CTIO $R$ filters,
Convolving the filter response with the MARCS models
and a Cardelli et al.\ (1989) reddening law confirms that this is a reasonable approximation even
for these very red stars, with coefficients of 0.75 (3400~K) to 0.82 (4300~K).  We adopt 0.81,
 which is
typical of our median temperatures.  We list the dereddened $(V-R)_0$ colors in 
Table~\ref{tab:bb}, along with the derived temperatures and bolometric luminosities.  For the
latter we assume the same $A_V$ as for the model fitting (i.e., Table~\ref{tab:results}), but
with the BCs derived from the $(V-R)_0$ effective temperatures.

In Fig.~\ref{fig:vmrteffs}(a) we show the comparison between the temperatures we derive from this
method compared to those from fitting the optical spectrophotometry.  
We see that there is very little difference.  There
is a slight offset in effective temperature compared to fitting the optical spectrophotometry (-30~K
for both the LMC and SMC), much less worrisome than the -105~K and -170 K found from the
$(V-K)_0$ fitting.  In Fig~\ref{fig:vmrteffs}(b) we show the comparison between the bolometric
luminosities.  These agree extremely well, as expected, given the only slight offset in
$T_{\rm eff}$.  We are forced to conclude that there is nothing wrong with our basic analysis
technique, and that at present $(V-K)_0$ gives slightly inconsistent answers, probably due to
the limitations of static 1D  models.

\section{Results}
\label{Sec-results}

\subsection{Reddening}
\label{Sec-reddening}

Paper I noted that many of the RSGs in Galactic OB associations show significantly
higher reddening than the early-type stars in the same clusters and associations; this matter was investigated more fully by Massey et al.\ (2005a), who demonstrated
that this extra reddening was likely due to circumstellar dust, and amounted to
as much as 4-5~mag of extra extinction at $V$ for stars with the highest
dust mass-loss rates and highest bolometric luminosities.  This was in accord
with a simple calculation of how much extinction one would {\it expect} just given
the observed dust mass-loss rates and reasonable assumptions. The criticism is easily leveled that of course the Galactic OB associations suffer from variable reddenings, and that such evidence is therefore somewhat dependent upon the sample selection.  We vowed to reexamine this issue in the Magellanic Clouds, where the extinction is generally low, and uniform (Massey et al.\ 1995, van den Bergh 2000).  Of course, the lower
metallicities of the Clouds should result in lower dust mass-loss rates.  We hope to
measure the dust mass-loss rates ourselves using the {\it Spitzer Space
Telescope (SST)},, 
but even if this rate scales linearly with metallicity, we
would expect to see RSGs with a considerable amount of extra extinction in the
Clouds.

In Fig.~\ref{fig:avs} we compare the $A_V$ values found from our model fitting of RSGs
to the distribution of reddenings for OB stars found by Massey et al.\ (1995).  Clearly the excess reddening we expected to find is in fact present.  The peaks are shifted by several tenths of a magnitude to higher
values, and there are a significant number of stars with substantially more reddening than that.  

\subsection{The H-R Diagram}
\label{Sec-evol}

With the improvements to the effective temperature scale given in Paper~I, we found excellent
agreement between the placement of RSGs in the H-R diagram and the locations of the
stellar evolutionary tracks from the Geneva group (i.e., Meynet \& Maeder 2003).
These tracks included new opacities, and spanned a range of initial rotational velocities (0-300 km s$^{-1}$). 

\begin{figure}
\epsscale{0.95}
\plotone{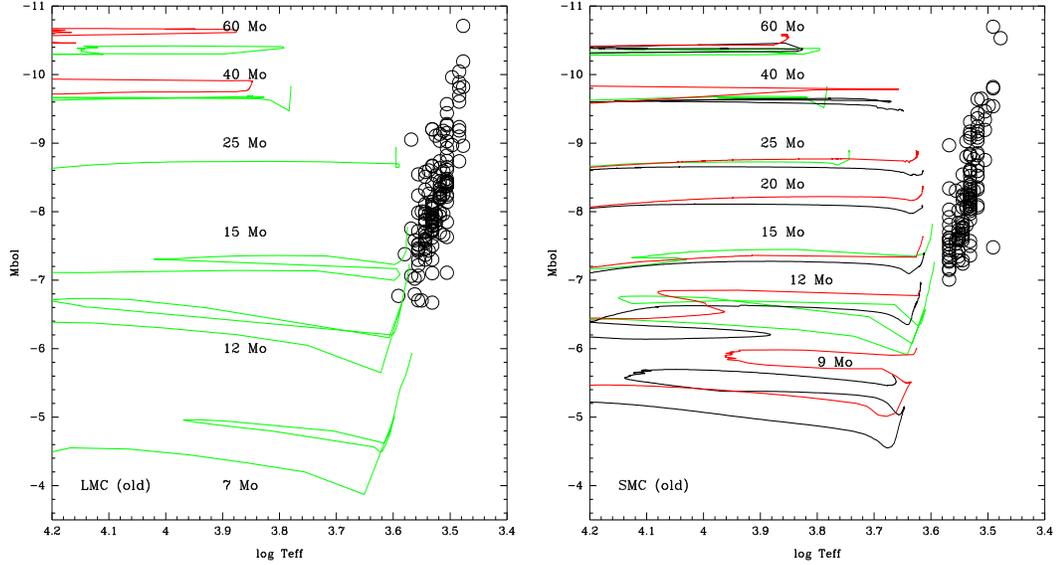}
\vskip -200pt

\caption{\label{fig:oldhrd}  The location of Magellanic Cloud RSGs (from the literature)
compared to the evolutionary tracks.
This figure has been adapted from Massey \& Olsen (2003), and shows the mismatch between
modern evolutionary theory and the canonical  locations of RSGs in the SMC and LMC, where
the effective temperatures and bolometric corrections are based on
those in the literature with a modest correction for metallicity
(see Massey \& Olsen 2003).  The tracks do not produce stars as
luminous and as cool as those ``observed".  A similar problem for
Galactic metallicity was solved by a significantly revised effective
temperature scale based upon fitting the MARCS stellar atmosphere
models to optical spectrophotometry.  The older, non-rotation
evolutionary tracks which include the effects of overshooting are
shown in green, and come from Schaerer et al.\ (1993)  for the LMC
and Charbonnel et al.\  (1993) for the SMC.  The newer evolutionary
tracks (when available) are shown in black (zero rotation) and in
red (300 km s$^{-1}$ initial rotation), and come from Meynet \&
Maeder (2005) for the LMC, and Maeder \& Meynet (2001) for the SMC.
}
\end{figure}

\begin{figure}
\epsscale{1.00}
\plotone{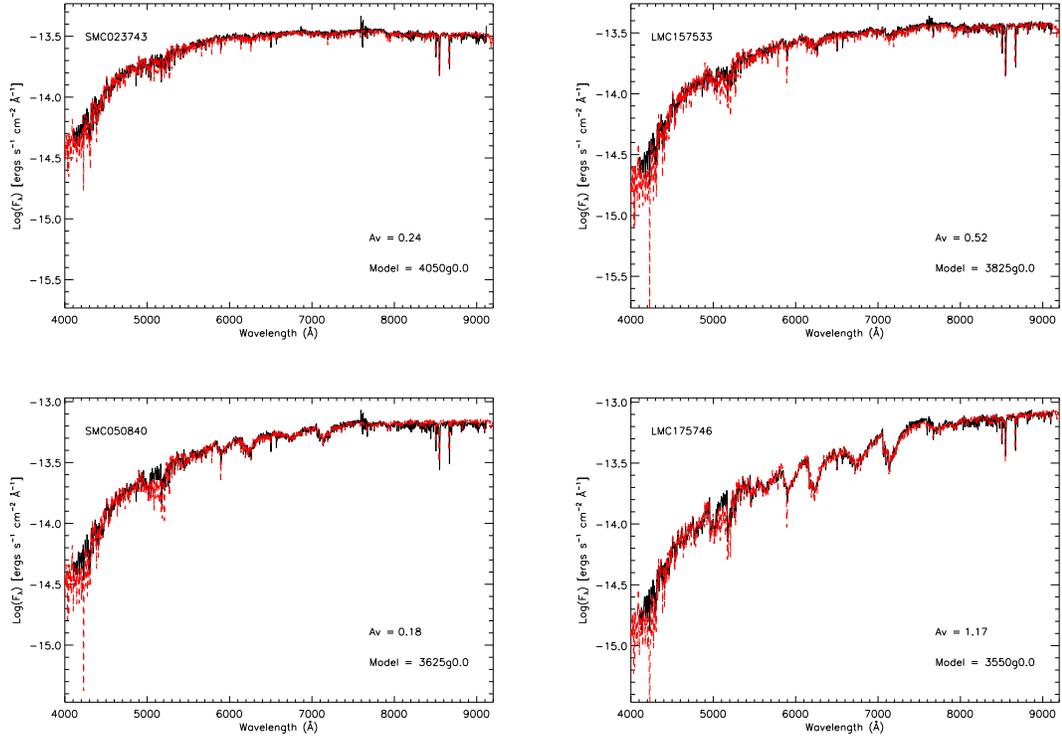}
\vskip -250pt
\caption{\label{fig:samples}  Model fits to optical spectrophotometry.  
The observed spectral energy
 distributions are shown in black, and the adopted
 model fits in red.  The models have been reddened
 by the indicated amount using the standard
 R$_V$=3.1 reddening law of Cardelli et al. (1989).
Four examples are shown here:
 SMC023743 (K2 I), LMC 157533 (K5 I),
SMC050840 (M1 I), and LMC 175746 (M3 I).  The individual
fits are shown as Figures 2.2-2.76 in the
electronic edition.
}
\end{figure}

\begin{figure}
\epsscale{1.0}
\plotone{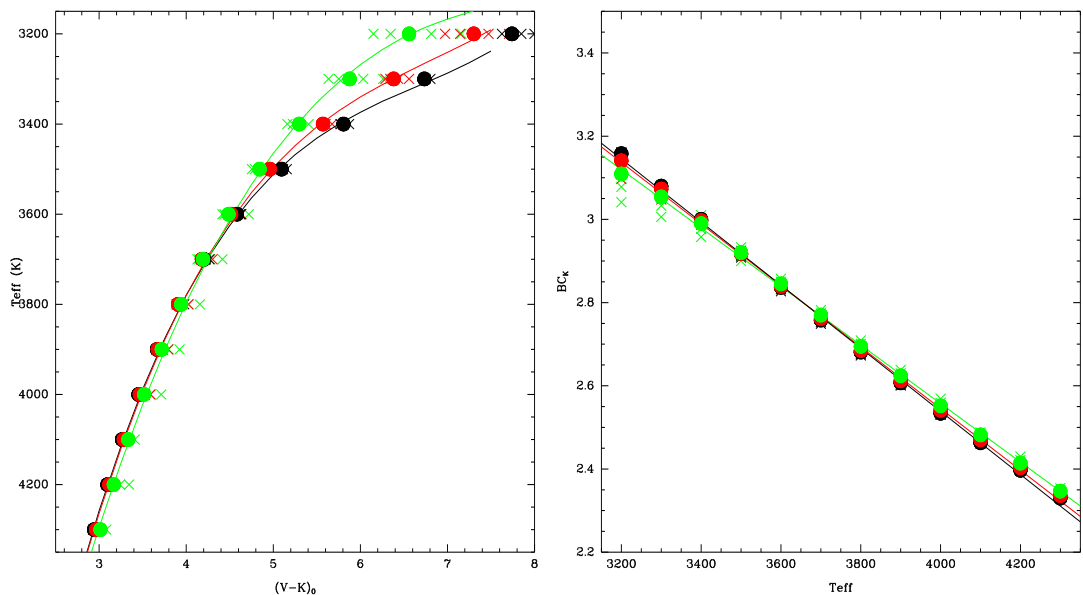}
\vskip -200pt
\caption{\label{fig:vmkfit}The transformation relations for K-band photometry.
Black represents the results from the MARCS models with solar metallicity,
while red represents the LMC and green the SMC.  The filled circles show
the values for $\log g=0.0$, while the x's cover the $\log g$ range from -1 to +1.
The solid curves denote the fits from \S~\ref{Sec-vmk}. (a) Transformation between
$T_{\rm eff}$ and $(V-K)_0$. (b) Transformation between BC$_K$ (the bolometric correction
at the K-band) and effective temperature.  
}
\end{figure}

\begin{figure}
\epsscale{0.68}
\plotone{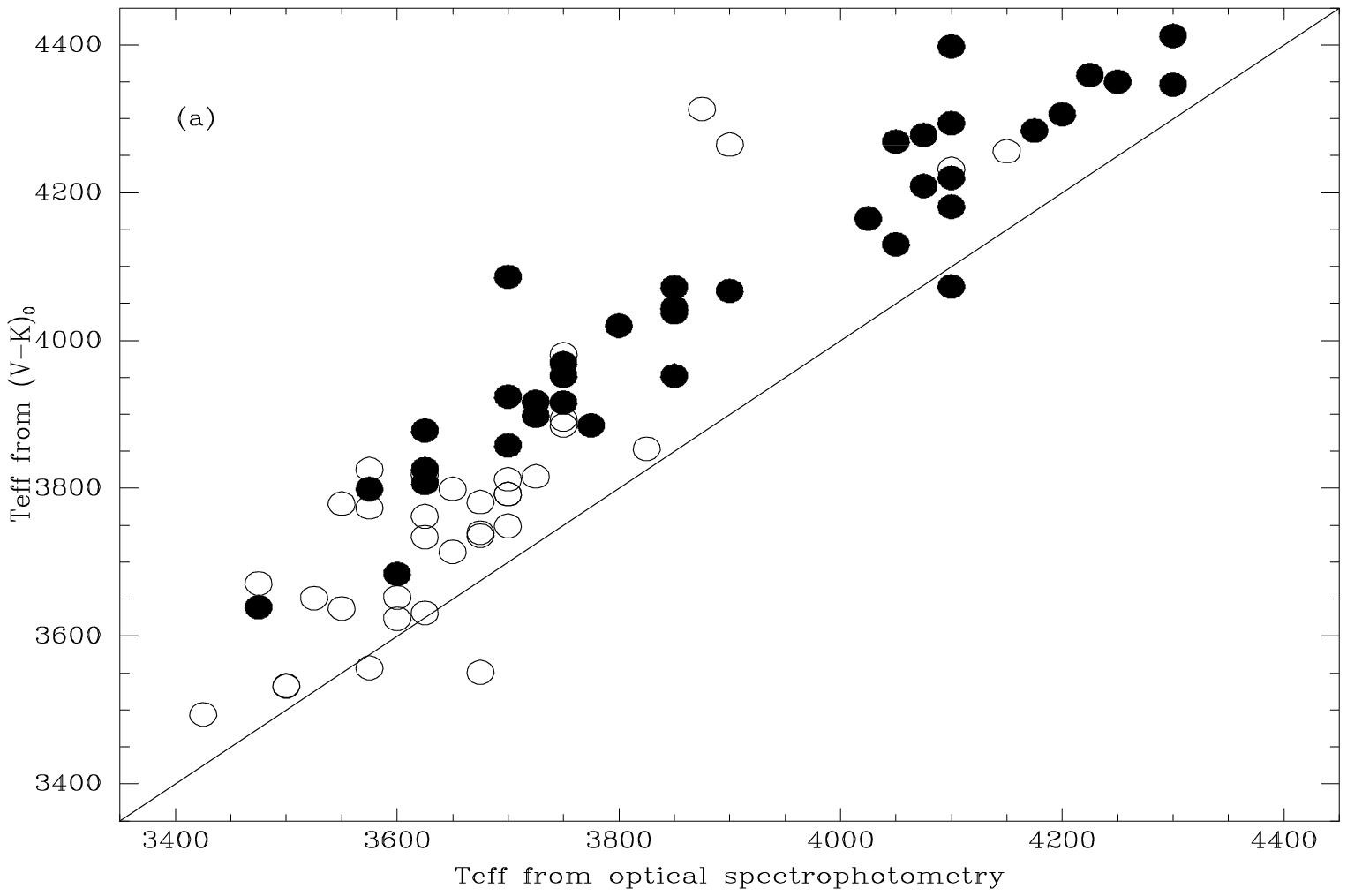}
\plotone{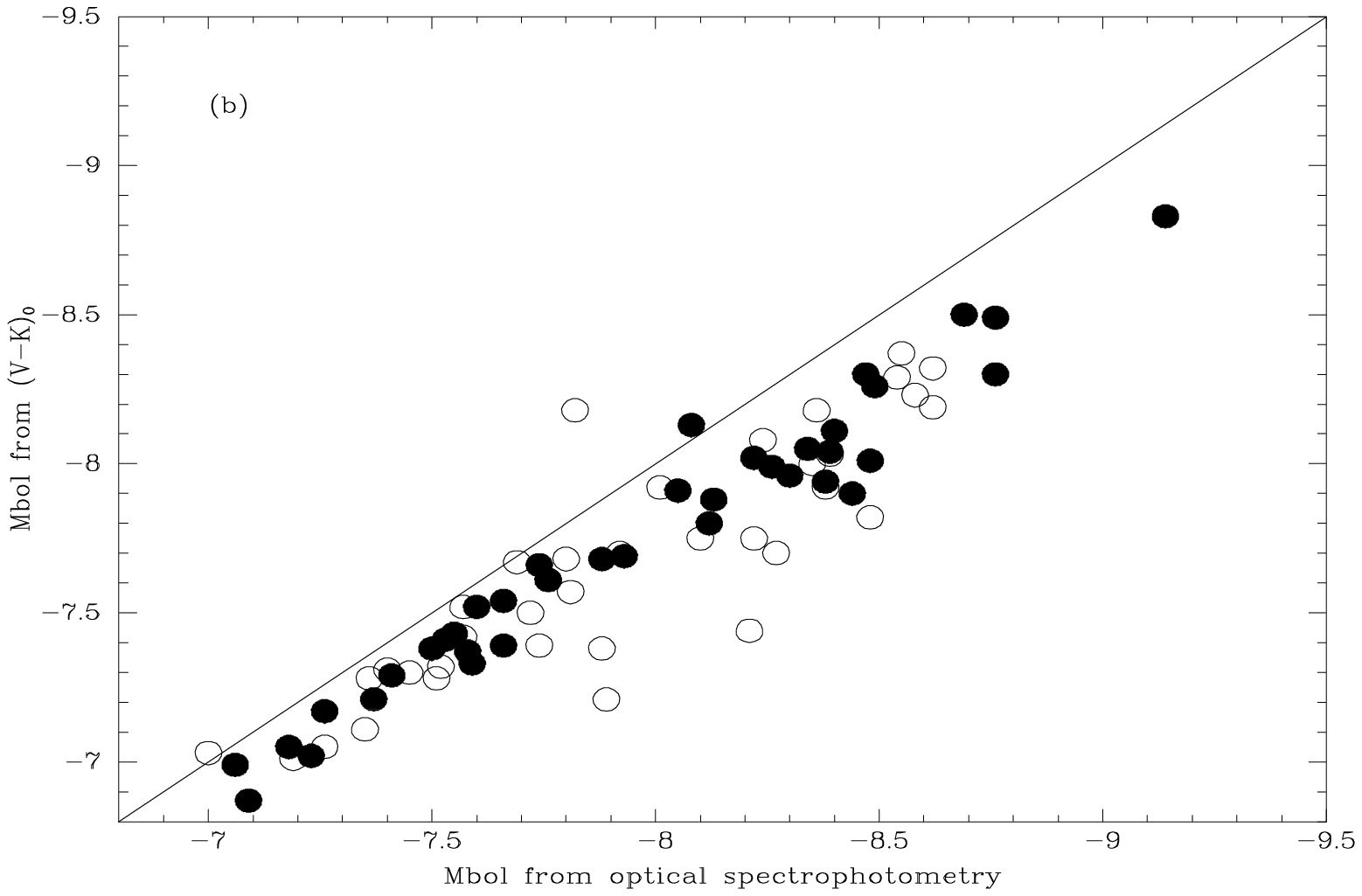}
\caption{\label{fig:diff} The physical properties derived from $(V-K)_0$ compared to
those obtained by fitting the optical spectrophotometry. 
 Filled circles show the data for the SMC, while open
circles show the results for the LMC.  The solid lines shows the 1:1 relation. (a) The effective
temperatures $T_{\rm eff}$ found from $(V-K)_0$ are systematically higher than those
obtained from the band depths of TiO when fitting the optical spectrophotometry. The median
offset is -105 K (LMC) and -170 K (SMC). (b) The bolometric magnitudes are similarly affected,
with a median offset of -0.23~mag (LMC) and -0.21~mag (SMC). 
}
\end{figure}

\begin{figure}
\epsscale{0.72}
\plotone{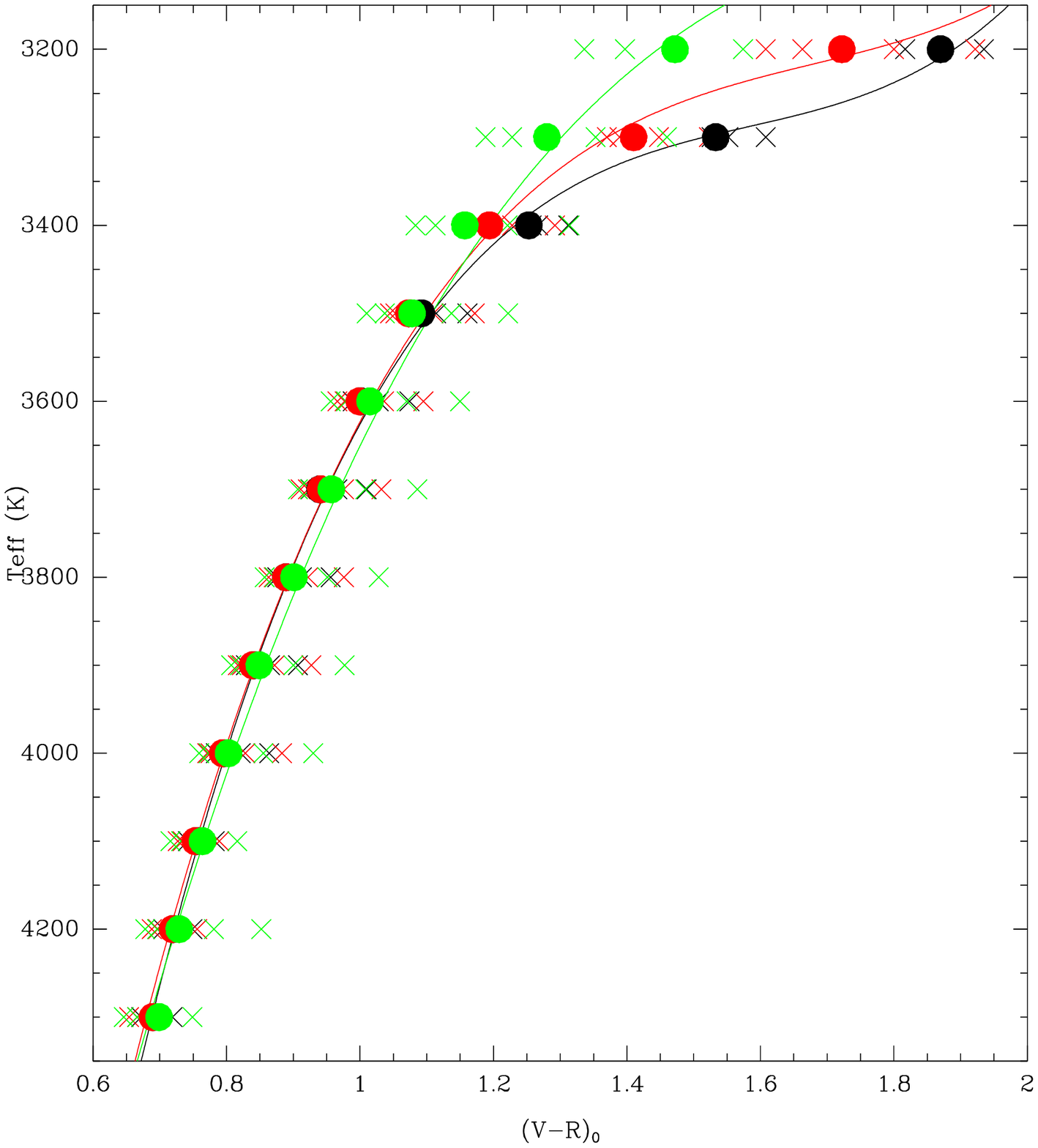}
\caption{\label{fig:vmrfit} The transformation relation for $(V-R)_0$ photometry.
Black denotes the Milky Way, red the LMC, and green the SMC.  The filled circles
show the values for $\log g=0.0$, while the x's cover the $\log g$ range from -1 to +1.
The fit was performed excluding the extremes in $\log g$, although these data are shown
here.
}
\end{figure}

\begin{figure}
\epsscale{0.68}
\plotone{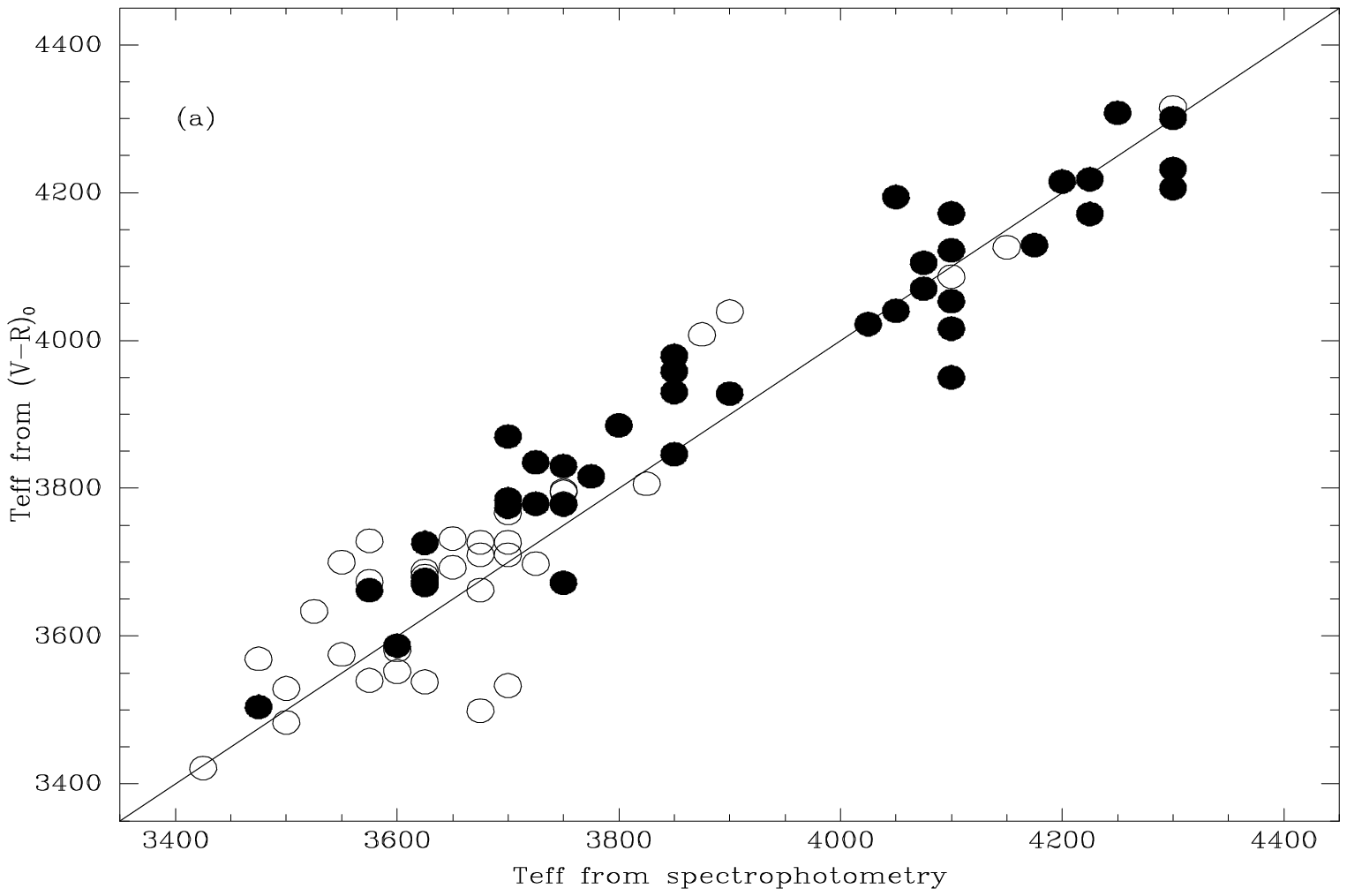}
\plotone{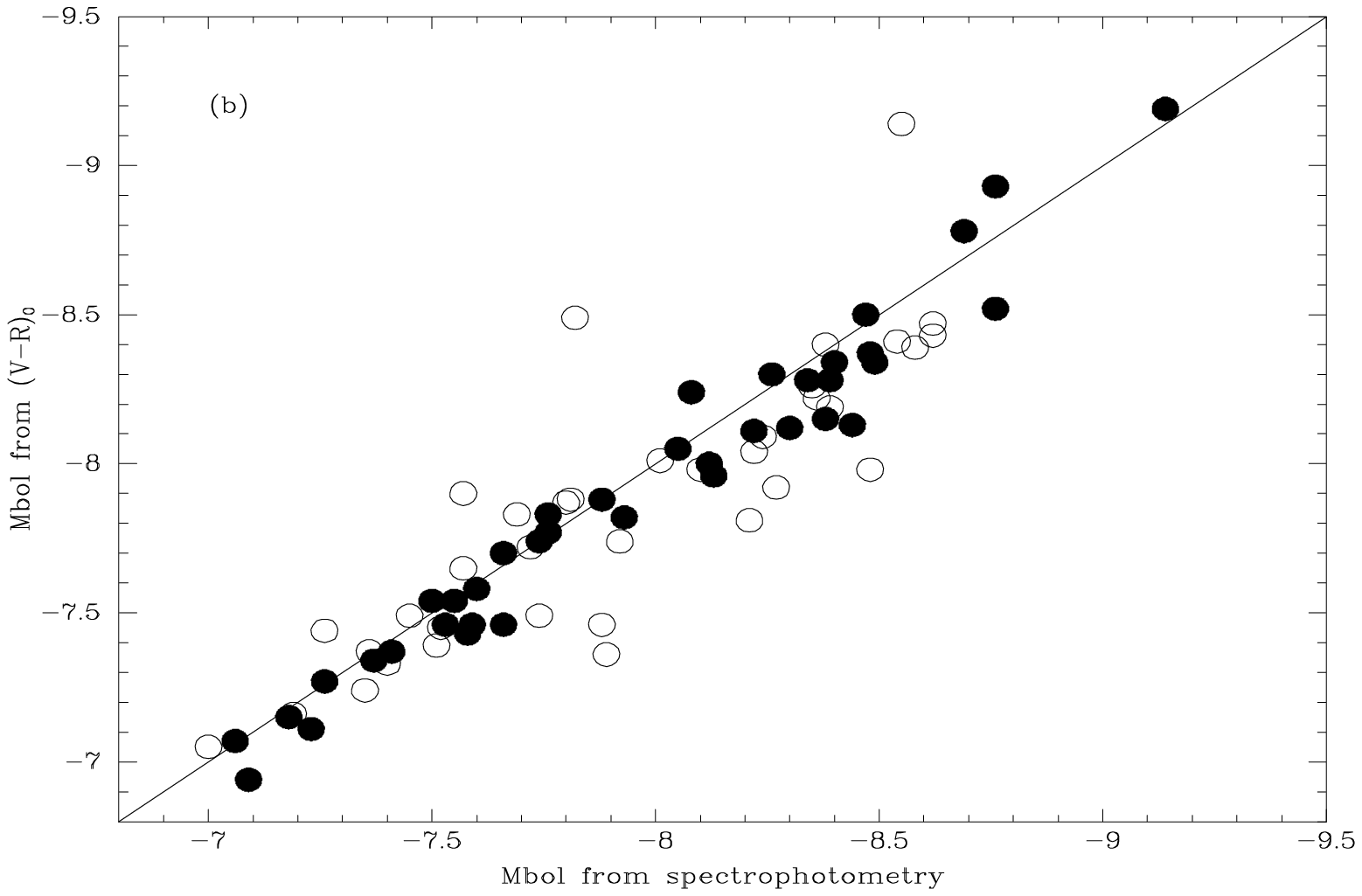}
\caption{\label{fig:vmrteffs} The physical properties derived from $(V-R)_0$ compared to
those obtained by fitting the optical spectrophotometry. 
 Filled circles show the data for the SMC, while open
circles show the results for the LMC.  The solid lines shows the 1:1 relation. (a) The effective
temperature $T_{\rm eff}$ found from $(V-R)_0$ show no systematic difference for the LMC
when compared to the values derived from optical spectrophotometry.  The data show only
a slight offset.  The formal median differences are -30~K for both galaxies. (b) The bolometric
luminosities computed using the $(V-R)_0$ colors to define the $T_{\rm eff}$ used for 
computing the bolometric corrections at $V$.  As expected, the differences are slight,
given the good agreement of the temperatures, with a median difference of -0.11 mag (LMC)
and -0.04 mag (SMC).
}
\end{figure}

\begin{figure}
\epsscale{0.48}
\plotone{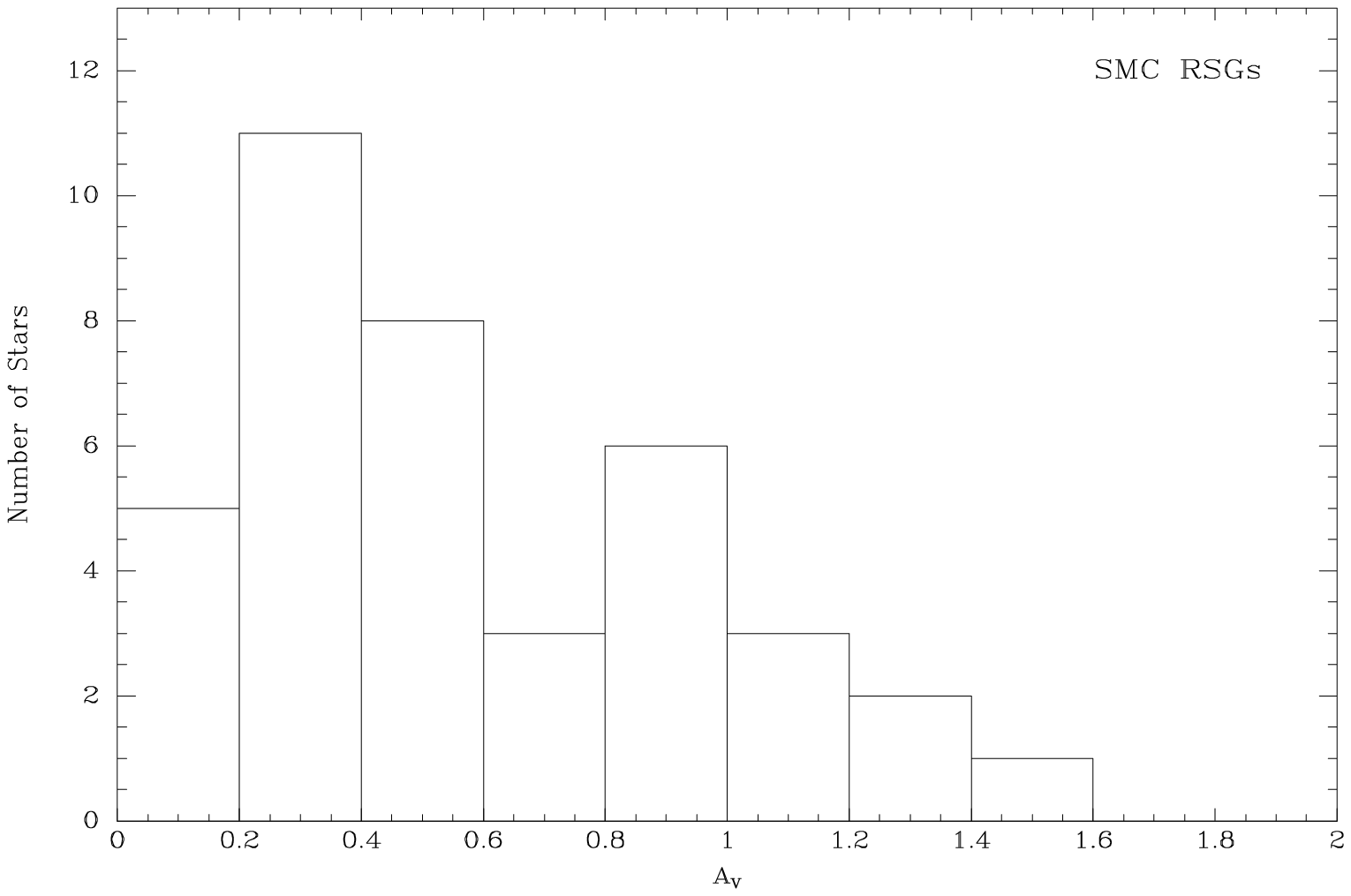}
\plotone{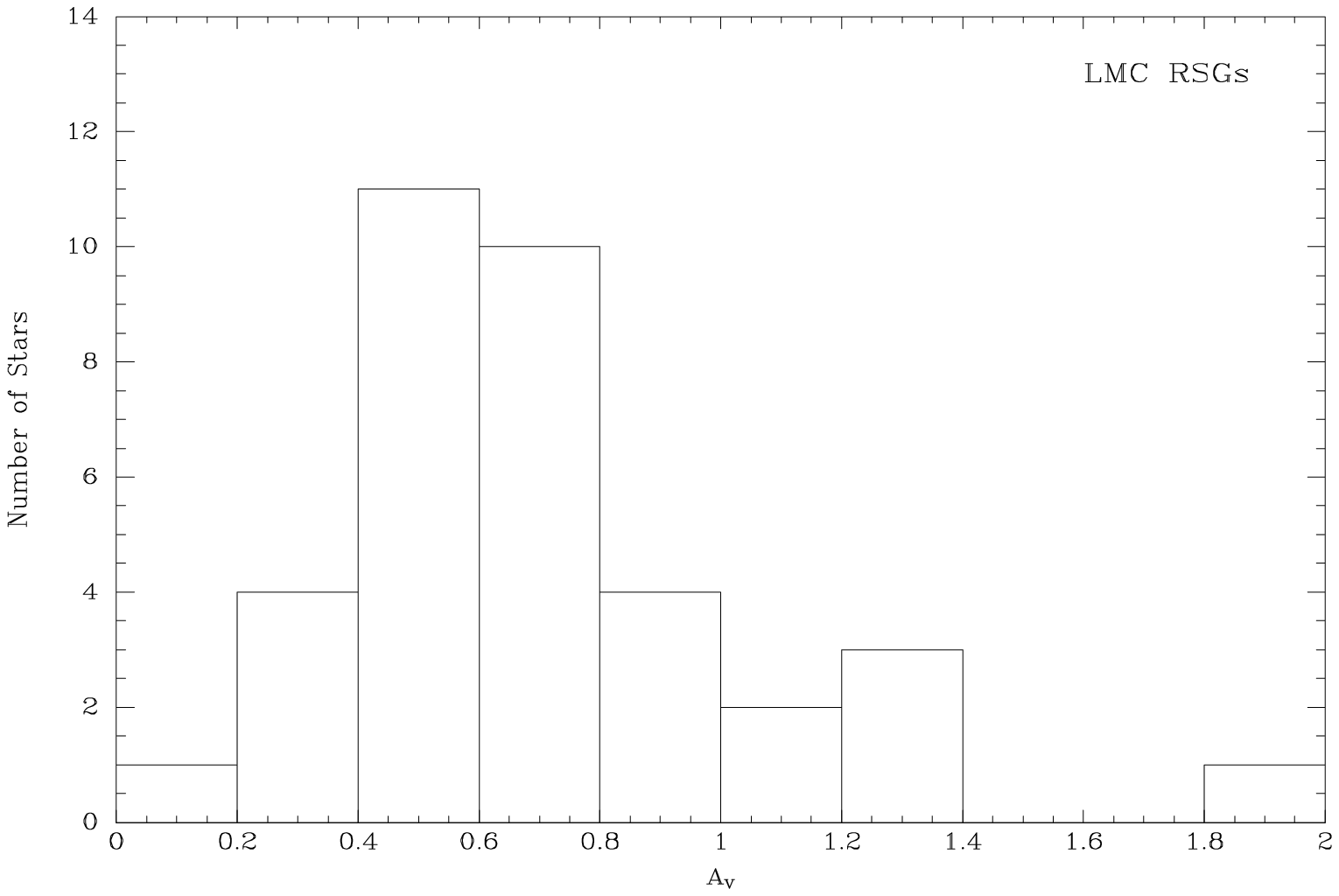}
\plotone{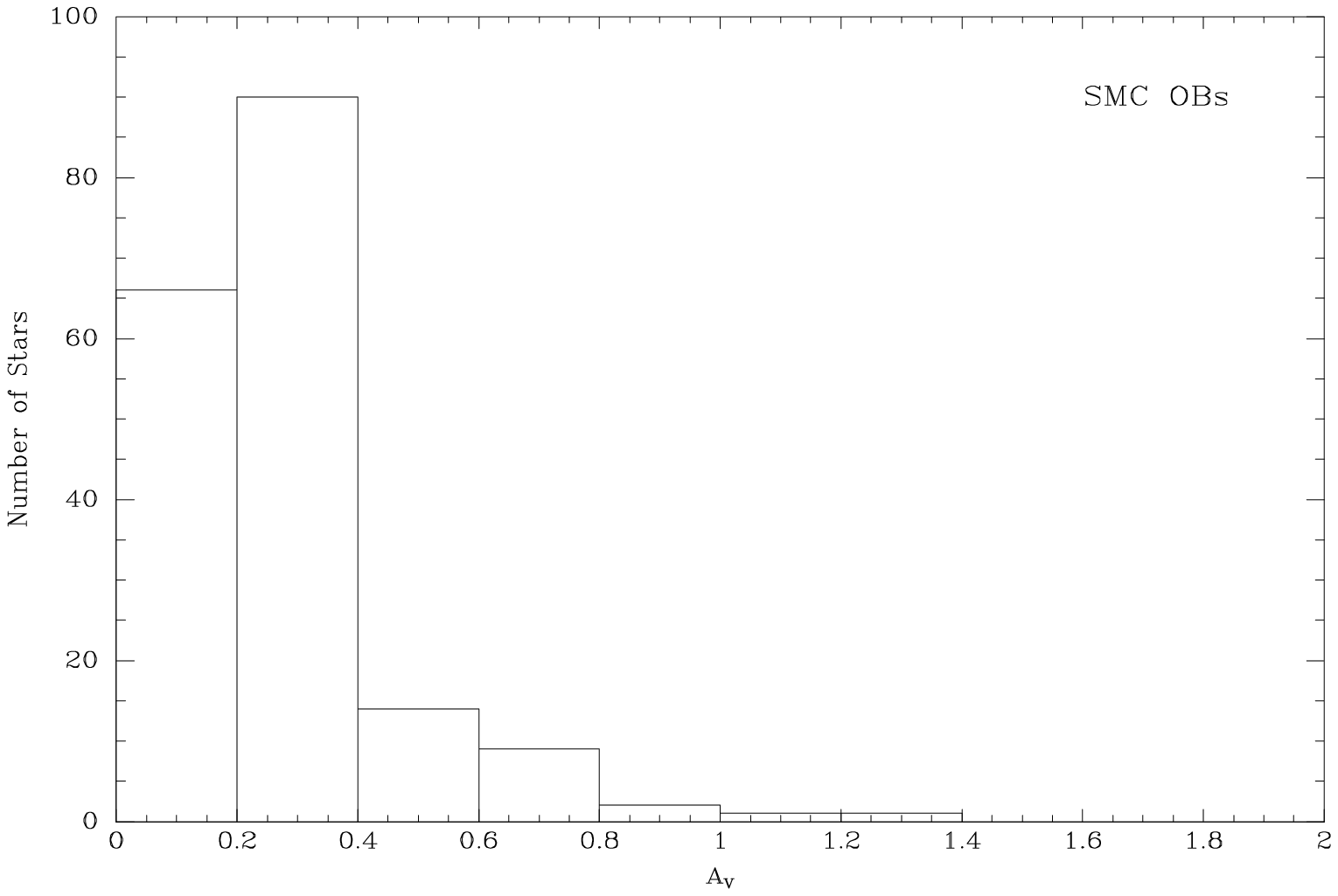}
\plotone{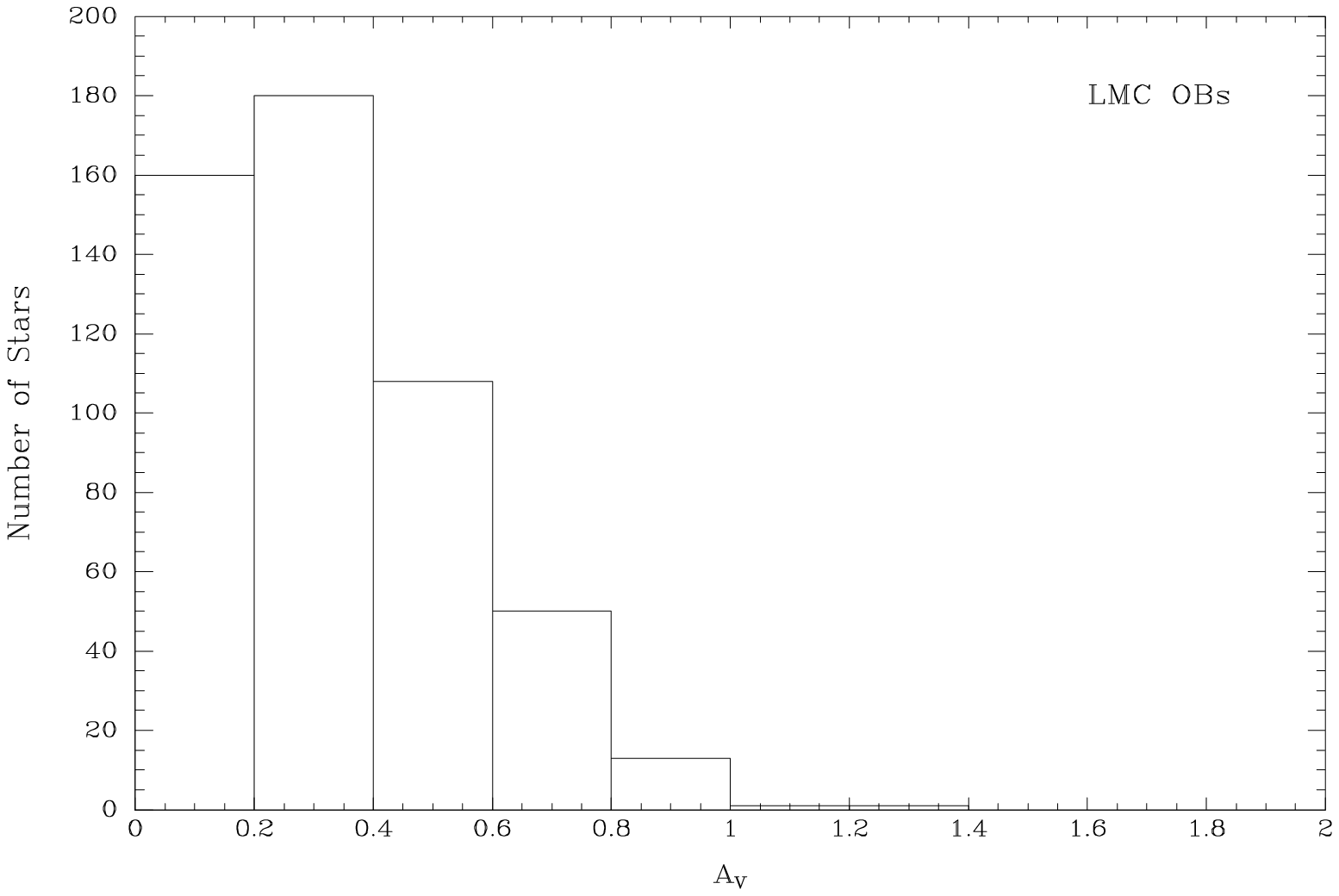}
\caption{\label{fig:avs} The distribution of $A_V$ of RSGs is compared to that of OB stars.
In both the SMC (left) and LMC (right) there is significantly more reddening for RSGs than is
seen for OB stars.  For the SMC the RSGs have $\bar{A}_V=0.60\pm0.06$, compared
to the OB stars, which have $\bar{A}_V=0.24\pm0.01$.  For the LMC, the RSGs have
$\bar{A}_V=0.73\pm0.06$ compared to the OB stars which have $\bar{A}_V=0.32\pm0.01$,
The errors quoted are the standard deviations
of the mean.  In addition, the distributions in $A_V$ are considerably broader for the RSGs,
as is obvious from the figure: the O stars have quite a narrow distribution in $A_V$,
with $\sigma=0.25$ for each Cloud, while the distribution for the RSGs is broader, 0.39~mag for
each Cloud.
This is consistent with our finding for the Milky Way (Paper~I and Massey et al.\ 2005a).}
\end{figure}

\begin{figure}
\epsscale{0.95}
\plotone{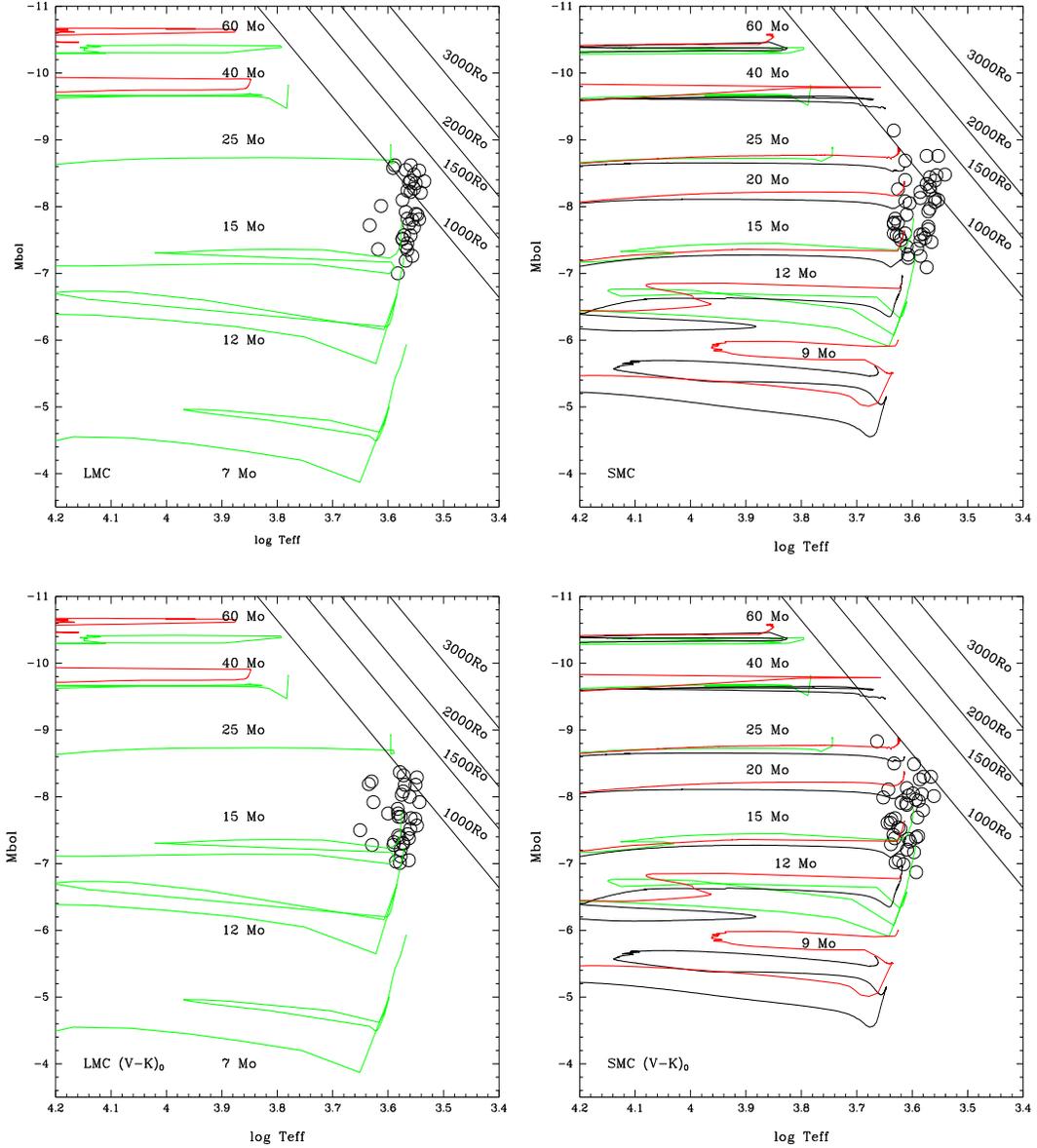}
\vskip -100pt
\caption{\label{fig:hrd} The location of Magellanic Cloud RSGs compared to the evolutionary tracks
(this work). In the top two panels we show the location of the RSGs in the H-R diagram for the
LMC (left) and SMC (right), where the effective temperatures and bolometric luminosities come from
fitting the MARCS models to the optical spectrophotometry.  Compare these to those shown in 
Fig.~\ref{fig:oldhrd}.  The same evolutionary tracks are plotted here
as in Fig.~\ref{fig:oldhrd}.
The bottom two panels show the same locations as derived from the $(V-K)_0$ calibration.
}
\end{figure}

In Fig.~\ref{fig:hrd}(a) and (b) we make a similar comparison now using the results from fitting
the optical spectrophotometry (i.e., Table~\ref{tab:results}).  ``Modern" models (including rotation
and the revised opacities) are available at present only for higher masses for the LMC
(Meynet \& Maeder 2005), but a full set for the SMC is available (Maeder \& Meynet
2002).  We complement these with the older Geneva models, as shown in green
(Schaerer et al.\  1993 for the LMC and Charbonnel et al.\  1993 for the SMC).  
 We see that for the LMC there is now excellent agreement between the tracks
 and the ``observed" (revised) locations of RSGs in the H-R diagram.
For the SMC the agreement is
poorer.  First,
the newer tracks do not extend to quite as low effective temperatures as do the data,
although we note that the older tracks (shown in green) do.   Still,
this discrepancy is small compared to the past (Fig.~\ref{fig:oldhrd}).  

More
interesting is the large spread in effective temperatures of RSGs of a given
luminosity in the SMC compared to that in the LMC.
This effect is just what is expected: Maeder \& Meynet (2001) found that, in part
because massive stars formed at low metallicity have relatively weak stellar
winds, little angular momentum is removed and hence rotational mixing is 
of increased importance at low metallicites.  
In this case, the helium content of red supergiants depends strongly on the 
rotational velocities during the main-sequence phase, and at higher helium
content the tracks stop at warmer temperatures.

We can compare these two H-R diagrams to what would be obtained if instead we relied upon
the $(V-K)_0$ calibration.  As discussed above (\S~\ref{Sec-vmk}) the $(V-K)_0$ calibration
produces somewhat warmer temperatures than does the fitting of the optical spectrophotometry
or those calculated from the broad-band $(V-R)_0$ colors.  The results are similar, as shown in 
Fig.~\ref{fig:hrd}(c) and (d).  No matter which calibration is used, the agreement with the
evolutionary tracks is satisfactory, and is a vast improvement over the situation shown in
Fig.~\ref{fig:oldhrd}.  And, in either case the dispersion for the SMC is considerably greater
than that for the LMC.

\subsection{Effective Temperatures Scales, the Hayashi Limit, and the
Distributions of Spectral Subtypes}

\label{Sec-met}

\begin{figure}
\epsscale{1.0}
\plotone{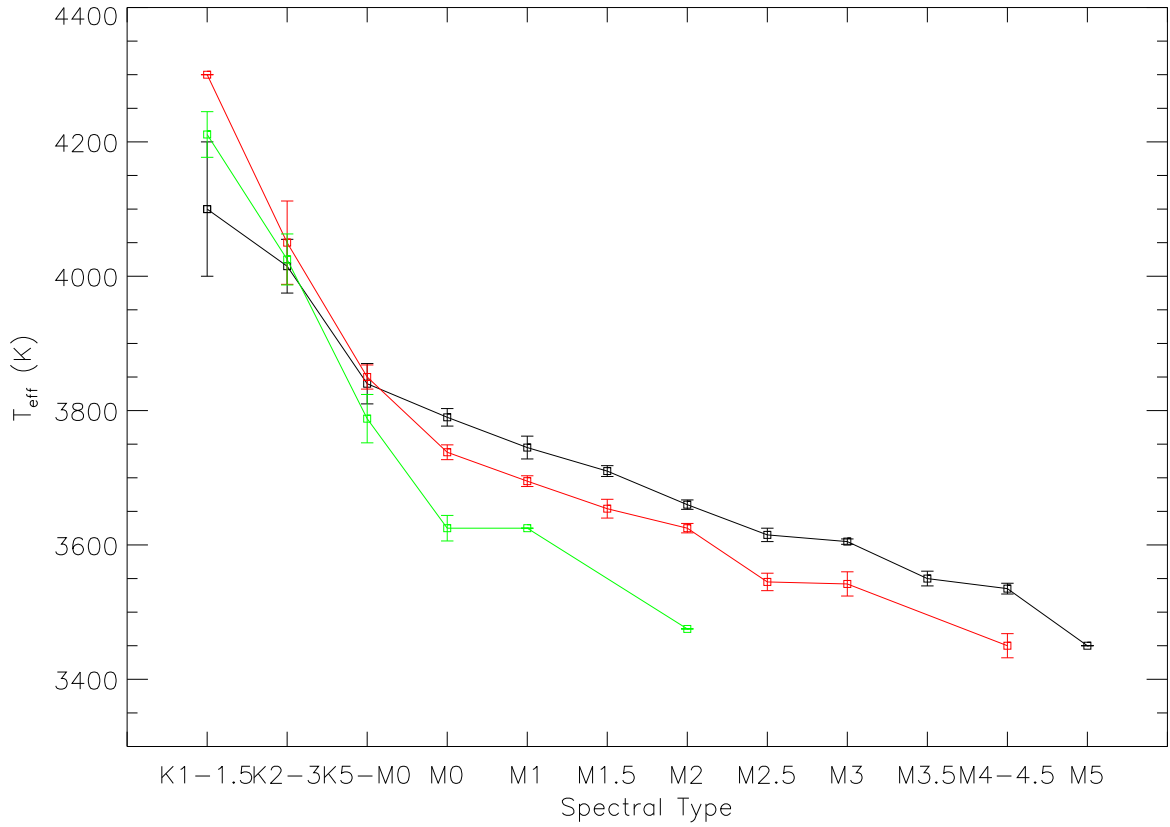}
\caption{\label{fig:tscale} Effective temperature scales for Galactic (black), 
LMC (red), and SMC (green) RSGs. The error bars reflect the standard deviation of the means
from Table~\ref{tab:tscale}.}
\end{figure}

The new effective temperature scales are compared in Table~\ref{tab:tscale},
and are shown in Fig.~\ref{fig:tscale}.    For the K supergiants, the scales
agree to within the errors.  For cooler stars,
the effective temperature scale
is about 50~K cooler for LMC M-type supergiants than Galactic M-type
supergiants.   It is about  
150~K cooler for SMC M-type supergiants  than Galactic M-type supergiants.

What we find is that while
an M2~I star in the Milky Way will have  $T_{\rm eff}$=3660~K,
a star of similar effective temperature in the LMC
would be spectroscopically identified as an M1.5~I, i.e., half a type earlier,
purely due to the effect that metallicity has on the depth of the TiO bands.
A star with the same effective temperature in the SMC would be spectroscopically
identified as an M0~I, i.e., two spectral subtypes earlier than in the
Milky Way.

This offers a partial, but incomplete, explanation for the shift in spectral subtypes
first found by Elias et al.\ (1985), and confirmed by the more complete 
data of Massey \& Olsen (2003).  Massey \& Olsen (2003) found that
the average spectral type of a RSG is M2~I in the Milky Way, M1~I in the LMC,
and K5-K7~I in the SMC.  The spectral classifications given here are
based upon better data, but the sample may be somewhat skewed towards later
types.  We find (from Table~\ref{tab:tscale}) that the average spectral type of stars
in our sample is M2~I (Milky Way), M1.5~I (LMC) and K3~I (SMC).
Thus,
the change in the spectral appearance due to the change in abundance of
TiO might be enough to explain the small shift in spectral type from the Milky
Way to the LMC, but it is {\it not} enough to explain the relative lack of red supergiants
in the SMC.  Instead, we must look  to the stellar evolutionary tracks.

\begin{figure}
\epsscale{1.0}
\plotone{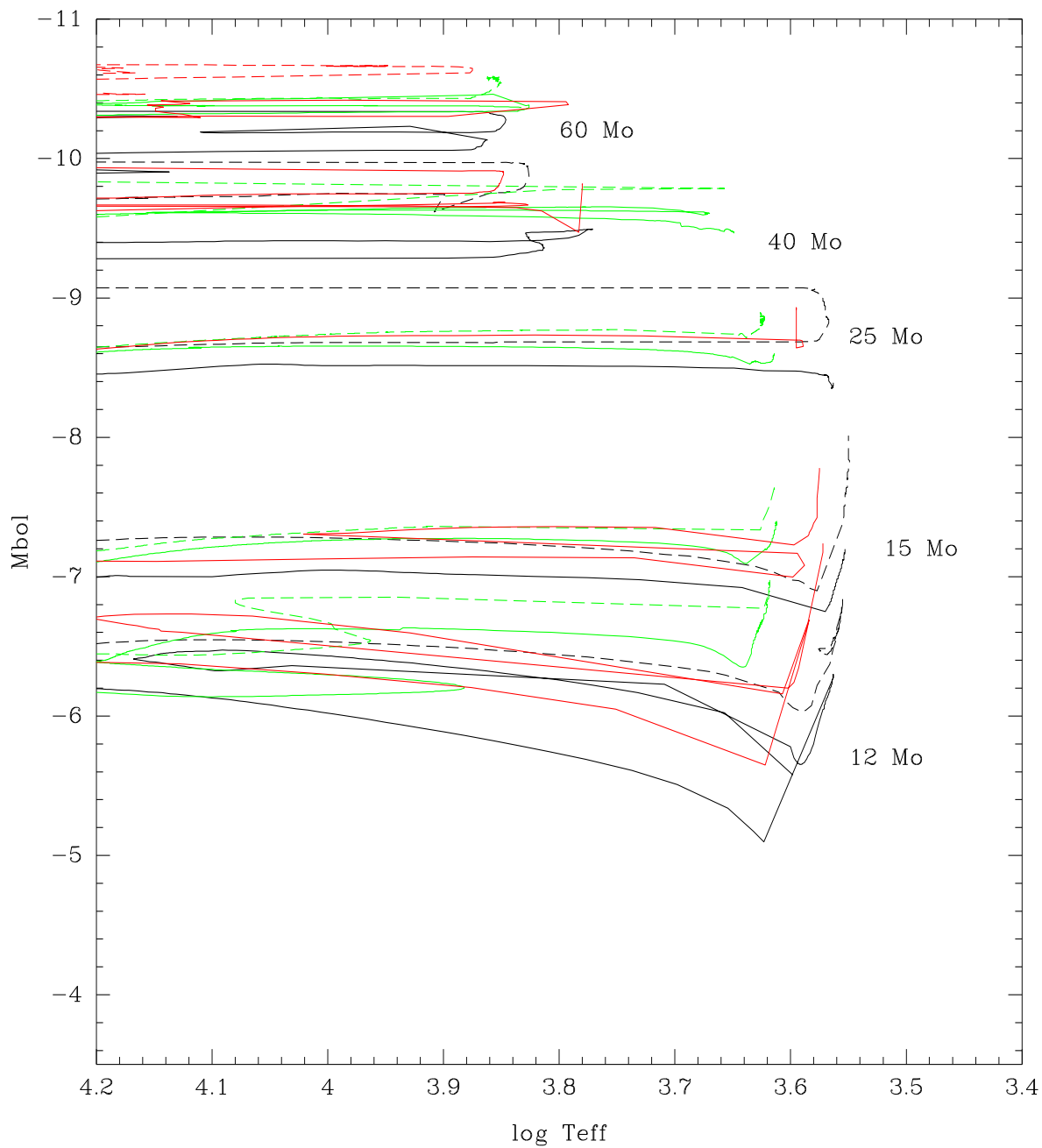}
\caption{\label{fig:compare} Comparison of evolutionary tracks at differing metallicities.  We show
the evolutionary tracks corresponding to the Milky Way (black, $z=0.02$), the LMC
(red, $z=0.008$, and SMC (green, $z=0.004$) from Meynet \& Maeder (2001, 2005),
Schaerer et al.\ (1993), and Charbonnel et al.\ (1993).   Solid curves indicate no
rotation, and dashed curves represent initial rotation velocities of 300 km s$^{-1}$. }
\end{figure}

Fig.~\ref{fig:compare} 
compares the evolutionary tracks for $z=0.02$ (Milky Way, shown in
black),
$z=0.008$ (LMC, shown in red), and $z=0.004$ (SMC, shown in green). 
We see that there is a clear shift of the coolest tip of the tracks
to warmer effective temperatures as the metallicity decreases.   Elias et al.\ (1985)
in fact attribute the shift of spectral subtypes to the effects of metallicity on the
location of the red supergiant  locus in the H-R diagram.  The
Hayashi limit denotes the largest radius a star of a given mass
can have and still be in
hydrodynamic equilibrium (Hayashi \& H\={o}shi 1961\footnote{Sugimoto \& 
Nomoto (1974) present a nice heuristic derivation based upon the argument that
the average density of a star must be greater than its photospheric density.}).
This line is nearly vertical in the H-R diagram, and shifts to warmer effective temperatures
at lower metallicities; we see the underlying physics reflected in the location of the coolest
extent of the evolutionary tracks.   Consistent with our results above (\S~\ref{Sec-evol}), this
shift {\it is} in accordance with what we need to explain the change in spectral subtype.  For 
a 15-25~$M_\odot$ star we expect a shift in effective temperatures of red supergiants
of about +500 K from Milky Way (black) to SMC (green).   This is actually
larger than the required shift, about 350 K (Table~\ref{tab:tscale}).  From the
Milky Way to the LMC the tracks shift by 100 to 150 K, in good agreement with
the shift in average spectral type from M2 (Milky Way) to M1-M1.5 (LMC).  
Thus we conclude that the shift in the spectral type due to the abundance of
TiO is a secondary effect, and that the main reason that the spectral subtypes are
earlier in the Clouds than in the Milky Way is due to the shift of the Hayashi limit
with metallicity. This is in accordance with the speculation offered by Elias et al.\ (1985).

\section{Summary and Conclusions}
\label{Sec-sum}

We have derived new physical properties of RSGs in the Magellanic Clouds, using the MARCS
stellar atmosphere models, following our treatment of Galactic RSGs in Paper I.  We find that
the effective temperatures of K supergiants are about the same in the SMC, LMC, and
Milky Way, but that the lower abundance of TiO leads to effective temperatures
that are about 50~K lower (LMC) and 150~K lower for M supergiants of the same
spectral subtype.
To put this in a more physical way, a star in the same place in the H-R diagram
that is called an M2~I in the Milky Way, would be of M1.5~I type in the LMC, and M0~I in the SMC.  This is not sufficient to explain the shift with metallicity
in the average spectral types between the Milky Way and the SMC, where the
average spectral types of RSGs change from M2~I to K3~I.  Instead, it is 
primarily the change in the Hayashi limit with metallicity that is responsible.
This agrees with the explanation offered by  Elias et al.\ (1985), who first observed
the shift in type.

Although the MARCS models give very good fits to the optical spectrophotometry, the results
derived from this fitting do result in temperatures that are systematically cooler than those that
would be derived from the observed $(V-K)_0$ colors.  The median
discrepancy is about -100~K for the Milky Way and the LMC, and -170~K for the SMC.  This systematic difference is likely due to the limitations of static 1D models.
The $(V-R)_0$ colors produce temperatures that are
more consistent with the optical spectrophotometry, with only a small offset (-30~K) for both
the SMC and the LMC.  

Although we would of course prefer for all techniques to give perfect agreement, we can
place this discrepancy into context.  A 175~K difference in the effective temperature
scale is a 4-5\% effect.    For comparison, recent revisions to the effective temperatures of
O-type stars have shifted the scale by 10\% (see discussion in Massey et al.\ 2005b).

The RSGs in our sample show higher extinction on average than do the OB stars in the
Clouds; this is consistent with our findings that many Galactic RSGs have higher extinction than
OB stars in the same clusters and associations (Paper I).  Massey et al.\ (2005a) argue that this
is a natural consequence of the fact that these stars are ``smokey", and produce circumstellar
dust.  Follow-up ground-based observations were successfully obtained
in December 2005 to derive the extinction properties of this dust, and we hope to extend this
work using the {\it SST.}

The newly derived properties provide an excellent match to stellar evolution tracks in the LMC,
but there is significantly more scatter for the SMC.  This may be due to the larger effect that rotational mixing  has in lower metallicity stars
(Maeder \& Meynet,  2001). The helium content of red supergiants
is changes significantly (typically by $\Delta Y = 0.10$ or more) as a function of
the rotation velocities during the main-sequence phase.  A higher helium abundance
results in the evolutionary tracks stopping at slightly warmer temperatures
than in the absence of such enrichment.
However, the amount of helium enrichment  critically depends
on the assumed physics of the models.
In this context, observational determinations of
the He/H ratios  would be very useful indeed.

Throughout this work we have made the approximation that the abundances of the SMC and LMC
scale by a single number, a fact which we know is not quite correct.  (For a good discussion of
this see Venn 1999 and Dufour 1984.)   As improved stellar models (with improved opacities and
which include the effects of rotation) become available at the relevant mass range (10-20 $M_\odot$)
we will better know what individual abundances to assume for  the model atmospheres.
In the meanwhile, we believe that the current study accurately represents the differential correction
to the derived properties of RSGs.  We also plan to extend this work to the metal-rich environment
found in the Andromeda Galaxy.

\acknowledgements
We are grateful to Hernan Tirado and Ricardo Venegas for excellent support with the RC spectrograph,
both during our observing and for the follow-up daytime tests to track down the elusive grating problem.
As always, observing at CTIO was a pleasure.  The preliminary part of this work was supported
under NSF grant AST 00-093060; additional support was provided through the Friends of Lowell
Observatory.
This paper made use of data products from the 2MASS, which is a joint project of the University of Massachusetts and the Infrared Processing and Analysis Center/California Institute of Technology, funded by the National Aeronautics and Space Administration and the National Science Foundation. An anonymous referee made useful comments on
the paper, leading to some substantial improvements; Deidre Hunter also kindly provided a critical
reading of the manuscript.

\begin{deluxetable}{l c c c c c  c  l  l}
\tabletypesize{\scriptsize}
\tablewidth{0pc}
\tablenum{1}
\tablecolumns{9}
\tablecaption{\label{tab:stars} Program Stars\tablenotemark{a}}
\tablehead{
\multicolumn{7}{c}{} &
\multicolumn{2}{c}{Spectral Type} \\ \cline{8-9}
\colhead{Star} &
\colhead{$\alpha_{\rm 2000}$} &
\colhead{$\delta_{\rm 2000}$} &
\colhead{$V$} &
\colhead{$B-V$} &
\colhead{$V-R$} & 
\colhead{$V-K_S$\tablenotemark{b}} &
\colhead{Old\tablenotemark{c}} &
\colhead{New} 
}
\startdata
SMC005092 &00 45 04.56&-73 05 27.4&12.90& 2.03& 1.18& 4.83&M1      I&M2       I  \\
SMC008930 &00 47 36.94&-73 04 44.3&12.68& 2.00& 1.06& 4.36&K7      I&M0       I  \\
SMC010889 &00 48 27.02&-73 12 12.3&12.20& 2.00& 1.06& 4.43&K7      I&M0       I  \\
SMC011101 &00 48 31.92&-73 07 44.4&13.54& 1.69& 0.99& 4.30&K7      I&K2.5     I  \\
SMC011709 &00 48 46.32&-73 28 20.7&12.43& 1.79& 0.94& 3.90&K7      I&K5-M0    I  \\
SMC011939 &00 48 51.83&-73 22 39.3&12.82& 1.81& 1.00& 4.21&M0      I&K2       I  \\
SMC012322 &00 49 00.32&-72 59 35.7&12.44& 1.93& 1.03& 4.16&M0      I&K5-M0    I  \\
SMC013740 &00 49 30.34&-73 26 49.9&13.47& 1.77& 0.96& 4.08&K7      I&K3       I  \\
SMC013951 &00 49 34.42&-73 14 09.9&13.00& 1.79& 0.93& 3.94&K7      I&K3       I  \\
SMC015510 &00 50 06.42&-73 28 11.1&12.59& 1.90& 0.95& 4.06&M0      I&K5       I  \\
SMC018136 &00 50 56.01&-72 15 05.7&11.98& 1.95& 1.01& 4.13&M0      I&M0       I  \\
SMC020133 &00 51 29.68&-73 10 44.3&12.33& 1.95& 1.03& 4.18&M0      I&M0       I  \\
SMC021362 &00 51 50.25&-72 05 57.2&12.89& 1.86& 0.95& 4.07&K5-M0   I&K5       I  \\
SMC021381 &00 51 50.46&-72 11 32.2&12.81& 1.81& 0.92& 3.81&K5-M0   I&K5       I  \\
SMC023401 &00 52 25.36&-72 25 13.3&12.99& 1.71& 0.84& 3.56&K5      I&K1       I  \\
SMC023743 &00 52 31.49&-72 11 37.3&12.98& 1.65& 0.84& 3.57&K5-M0   I&K2       I  \\
SMC025879 &00 53 08.87&-72 29 38.6&11.91& 1.77& 0.88& 3.46&K7      I&M0       I  \\
SMC030135 &00 54 26.90&-72 52 59.4&12.84& 1.68& 0.78& 3.35&K0-2    I&K2       I  \\
SMC030616 &00 54 35.90&-72 34 14.3&12.22& 1.85& 0.92& 3.88&K7      I&K2       I  \\
SMC034158 &00 55 36.58&-72 36 23.6&12.79& 1.78& 0.95& 3.88&K7      I&K2       I  \\
SMC035445 &00 55 58.84&-73 20 41.4&12.74& 1.77& 0.91& 3.76&M0      I&K1       I  \\
SMC042438 &00 58 08.71&-72 19 26.7&13.20& 1.59& 0.87& 3.84&K3-5    I&K2       I  \\
SMC043219 &00 58 23.30&-72 48 40.7&13.06& 1.84& 0.94& 3.95&M0      I&K2       I  \\
SMC045378 &00 59 07.16&-72 13 08.6&12.93& 1.56& 0.92& 3.93&K5      I&K3       I  \\
SMC046497 &00 59 31.33&-72 15 46.4&12.40& 1.98& 0.99& 4.09&M1      I&K5-M0    I  \\
SMC046662 &00 59 35.04&-72 04 06.2&12.90& 1.88& 1.07& 4.55&M2      I&K3       I  \\
SMC048122 &01 00 09.42&-72 08 44.5&12.19& 1.78& 0.89& 3.46&K3      I&K1       I  \\
SMC049478 &01 00 41.56&-72 10 37.0&12.17& 1.81& 0.99& 4.21&M0      I&K5-M0    I  \\
SMC050028 &01 00 55.12&-71 37 52.6&11.81& 1.82& 0.97& 3.77&M0      I&K1       I  \\
SMC050840 &01 01 15.99&-72 13 10.0&12.57& 1.95& 1.02& 4.20&M1-2    I&M1       I  \\
SMC054708 &01 02 51.37&-72 24 15.5&12.82& 1.81& 0.91& 3.74&K0      I&K1       I  \\
SMC055681 &01 03 12.98&-72 09 26.5&12.52& 1.65& 0.96& 3.93&M3      I&K5-M0    I  \\
SMC056732 &01 03 34.30&-72 06 05.8&12.86& 1.53& 0.94& 4.00&K7      I&K5-M0    I  \\
SMC057386 &01 03 47.35&-72 01 16.0&12.71& 1.57& 0.85& 3.74&K3-5    I&K1       I  \\
SMC057472 &01 03 48.89&-72 02 12.7&12.80& 1.83& 0.88& 3.83&K5-7    I&K2       I  \\
SMC059803 &01 04 38.16&-72 01 27.2&11.98& 1.95& 0.98& 3.88&M0-1    I&K2-3     I  \\
SMC060447 &01 04 53.05&-72 47 48.5&13.09& 1.64& 0.94& 3.91&M0      I&K2       I  \\
SMC067509 &01 08 13.34&-72 00 02.9&12.74& 1.68& 0.86& 3.57&K2      I&K1       I  \\
SMC069886 &01 09 38.08&-73 20 01.9&11.74& 1.95& 1.04& 3.95&M2      I&K5-M0    I  \\
LMC054365 &05 02 09.57&-70 25 02.4&13.26& 1.85& 1.10& 4.94&M3      I&M2.5     I  \\
LMC061753 &05 03 59.77&-69 38 15.0&13.16& 2.07& 1.16& 5.14&M2      I&M2       I  \\
LMC062090 &05 04 05.10&-70 22 46.7&12.50& 1.96& 1.00& 4.39&M1      I&M1       I  \\
LMC064048 &05 04 41.79&-70 42 37.2&13.28& 1.89& 1.19& 5.25&M3      I&M2.5     I  \\
LMC065558 &05 05 10.03&-70 40 03.2&12.62& 1.89& 1.01& 4.24&M0      I&M1       I  \\
LMC067982 &05 05 56.61&-70 35 24.0&12.76& 1.93& 1.09& 4.65&M4.5    I&M2.5     I  \\
LMC068098 &05 05 58.92&-70 29 14.6&13.11& 1.90& 1.04& 4.64&M1      I&M1.5     I  \\
LMC068125 &05 05 59.56&-70 48 11.4&13.43& 1.83& 1.20& 5.12&M4      I&M4       I  \\
LMC109106 &05 17 56.51&-69 40 25.4&12.96& 1.85& 1.02& 4.39&M2      I&M2.5     I  \\
LMC116895 &05 19 53.34&-69 27 33.4&12.43& 1.92& 1.03& 4.21&M3      I&M0       I  \\
LMC119219 &05 20 23.69&-69 33 27.3&12.14& 2.04& 0.98& 4.16&M3      I&M3       I  \\
LMC131735 &05 23 34.09&-69 19 07.0&12.65& 1.84& 0.89& 3.75&K7      I&K2       I  \\
LMC134383 &05 25 44.95&-69 04 48.9&13.46& 1.65& 1.21& 5.47&M3      I&M2.5     I  \\
LMC135720 &05 26 27.52&-69 10 55.5&13.57& 1.85& 1.35& 5.86&M3      I&M4.5     I  \\
LMC136042 &05 26 34.92&-68 51 40.1&12.24& 1.08& 1.09& 4.97&M1      I&M3       I  \\
LMC137624 &05 27 10.38&-69 16 17.6&13.16& 1.88& 1.02& 4.38&M0      I&M0       I  \\
LMC137818 &05 27 14.33&-69 11 10.7&13.33& 1.74& 1.20& 5.14&M3      I&M2       I  \\
LMC138405 &05 27 26.86&-69 00 02.0&13.08& 1.83& 1.02& 4.41&M0      I&M1       I  \\
LMC140296 &05 28 06.11&-69 07 13.5&13.12& 1.87& 1.18& 4.97&M1-2    I&M2       I  \\
LMC141430 &05 28 28.98&-68 07 07.8&12.30& 2.15& 1.24& 4.82&M0 I\tablenotemark{d}    &M1       I  \\
LMC142202 &05 28 45.59&-68 58 02.3&12.15& 1.65& 1.03& 4.60&M0-M1   I&M1.5     I  \\
LMC142907 &05 29 00.86&-68 46 33.6&13.05& 1.89& 1.06& 4.61&M1      I&M2       I  \\
LMC143877 &05 29 21.10&-68 47 31.5&11.82& 1.94& 0.95& 3.85&K7      I&K3       I  \\
LMC146126 &05 30 02.36&-67 02 45.0&11.17& 1.80& 0.84& 3.20&K5      I&K5       I  \\
LMC147199 &05 30 21.00&-67 20 05.7&12.73& 1.57& 1.20& 5.28&M4      I&M1.5     I  \\
LMC149721 &05 31 03.50&-69 05 40.0&12.71& 1.86& 0.97& 4.13&K5-7    I&M0       I  \\
LMC157533 &05 33 29.67&-67 31 38.0&13.16& 1.50& 0.99& 4.34&K5      I&K5       I  \\
LMC158317 &05 33 44.60&-67 24 16.9&13.35& 1.96& 1.12& 4.86&M2      I&M1       I  \\
LMC159974 &05 34 21.49&-69 21 59.8&12.72& 1.77& 0.91& 3.83&K2-5    I&K1       I  \\
LMC169754 &05 37 58.77&-69 14 23.7&13.21& 2.15& 1.13& 4.83&K2-3    I&K2       I  \\
LMC174714 &05 40 24.48&-69 21 16.6&13.13& 1.98& 1.21& 5.28&M4-5    I&M1.5     I  \\
LMC175464 &05 40 55.36&-69 23 25.0&12.90& 2.20& 1.22& 5.36&M2-3    I&M2       I  \\
LMC175746 &05 41 06.94&-69 17 14.8&13.30& 2.06& 1.26& 5.53&M3      I&M3       I  \\
LMC176890 &05 41 50.26&-69 21 15.7&12.85& 1.97& 1.01& 4.29&K7      I&M0       I  \\
LMC177150 &05 42 00.84&-69 11 37.0&13.80& 1.89& 1.20& 5.12&M1      I&M1.5     I  \\
LMC177997 &05 42 35.48&-69 08 48.3&12.56& 2.02& 1.08& 4.85&M2 I\tablenotemark{d}     &M1.5     I  \\
\enddata
\tablenotetext{a}{Star identification and optical photometry from Massey 2002}
\tablenotetext{b}{$K_S$ photometry from 2MASS point-source catalog.}
\tablenotetext{c}{Old spectral type from Massey \& Olsen 2003 for all stars unless otherwise noted.}
\tablenotetext{d}{Old spectral type from Elias et al.\ 1985}
\end{deluxetable}

\begin{deluxetable}{l l c c r r r c c}
\tabletypesize{\scriptsize}
\tablewidth{0pc}
\tablenum{2}
\tablecolumns{9}
\tablecaption{\label{tab:results} Results of Spectral Fits}
\tablehead{
\colhead{} &
\colhead{Spectral}&
\multicolumn{2}{c}{} &
\multicolumn{2}{c}{$\log g$} \\ \cline{5-6}
\colhead{Star} &
\colhead{Type} &
\colhead{$T_{\rm eff}$} &
\colhead{$A_V$} &
\colhead{Model} &
\colhead{Actual} &
\colhead{$R/R_\odot$} &
\colhead{$M_V$}  &
\colhead{$M_{\rm bol}$} 
}
\startdata
SMC005092 &M2      I&3475&0.40&-0.5&-0.4&1220& -6.40& -8.48\\
SMC008930 &M0      I&3625&0.56& 0.0&-0.3&1070& -6.78& -8.38\\
SMC010889 &M0      I&3600&0.09&-0.5&-0.3&1130& -6.79& -8.47\\
SMC011101 &K2.5    I&4200&1.43& 0.0& 0.2& 540& -6.79& -7.55\\
SMC011709 &K5-M0   I&3725&0.09& 0.0&-0.1& 830& -6.56& -7.93\\
SMC011939 &K2      I&4025&1.05& 0.0& 0.0& 750& -7.13& -8.05\\
SMC012322 &K5-M0   I&3750&0.56& 0.0&-0.2& 980& -7.02& -8.34\\
SMC013740 &K3      I&3750&0.34& 0.0& 0.2& 550& -5.77& -7.09\\
SMC013951 &K3      I&4225&1.12& 0.0& 0.2& 590& -7.02& -7.76\\
SMC015510 &K5      I&3850&0.68& 0.0&-0.1& 850& -6.99& -8.13\\
SMC018136 &M0      I&3575&0.09&-0.5&-0.4&1310& -7.01& -8.76\\
SMC020133 &M0      I&3625&0.22&-0.5&-0.3&1080& -6.79& -8.39\\
SMC021362 &K5      I&3775&0.25& 0.0& 0.0& 670& -6.26& -7.53\\
SMC021381 &K5      I&3800&0.28& 0.0& 0.0& 680& -6.37& -7.59\\
SMC023401 &K1      I&4075&0.40& 0.0& 0.3& 490& -6.31& -7.18\\
SMC023743 &K2      I&4050&0.25& 0.0& 0.3& 470& -6.17& -7.06\\
SMC025879 &M0      I&3700&0.03&-0.5&-0.3&1060& -7.02& -8.44\\
SMC030135 &K2      I&4050&0.28& 0.0& 0.2& 510& -6.34& -7.23\\
SMC030616 &K2      I&3850&0.40& 0.0&-0.1& 880& -7.08& -8.22\\
SMC034158 &K2      I&4075&0.90& 0.0& 0.1& 670& -7.01& -7.88\\
SMC035445 &K1      I&4100&0.65& 0.0& 0.1& 600& -6.81& -7.66\\
SMC042438 &K2      I&4250&0.99& 0.0& 0.3& 500& -6.69& -7.41\\
SMC043219 &K2      I&3850&0.28& 0.0& 0.2& 570& -6.12& -7.26\\
SMC045378 &K3      I&3850&0.47& 0.0& 0.1& 660& -6.43& -7.58\\
SMC046497 &K5-M0   I&3700&0.37& 0.0&-0.2& 990& -6.87& -8.30\\
SMC046662 &K3      I&4100&1.24& 0.0& 0.0& 730& -7.24& -8.08\\
SMC048122 &K1      I&4225&0.81& 0.0& 0.0& 740& -7.52& -8.26\\
SMC049478 &K5-M0   I&3700&0.34& 0.0&-0.3&1080& -7.07& -8.49\\
SMC050028 &K1      I&4300&1.36& 0.0&-0.2&1080& -8.45& -9.14\\
SMC050840 &M1      I&3625&0.19& 0.0&-0.2& 950& -6.52& -8.12\\
SMC054708 &K1      I&4300&0.99& 0.0& 0.2& 570& -7.07& -7.76\\
SMC055681 &K5-M0   I&4100&1.18& 0.0&-0.1& 850& -7.56& -8.40\\
SMC056732 &K5-M0   I&3725&0.25& 0.0& 0.0& 730& -6.29& -7.66\\
SMC057386 &K1      I&4300&0.87& 0.0& 0.2& 570& -7.06& -7.74\\
SMC057472 &K2      I&4100&0.65& 0.0& 0.2& 580& -6.75& -7.60\\
SMC059803 &K2-3    I&4100&0.93& 0.0&-0.2& 970& -7.85& -8.69\\
SMC060447 &K2      I&3900&0.50& 0.0& 0.1& 580& -6.31& -7.37\\
SMC067509 &K1      I&4175&0.56& 0.0& 0.2& 540& -6.72& -7.50\\
SMC069886 &K5-M0   I&3750&0.28&-0.5&-0.3&1190& -7.44& -8.76\\
LMC054365 &M2.5    I&3525&0.56& 0.0&-0.2& 900& -5.80& -7.88\\
LMC061753 &M2      I&3600&0.68& 0.0&-0.1& 830& -6.02& -7.80\\
LMC062090 &M1      I&3700&0.47& 0.0&-0.1& 830& -6.47& -7.92\\
LMC064048 &M2.5    I&3500&0.40& 0.0&-0.2& 880& -5.62& -7.81\\
LMC065558 &M1      I&3725&0.31& 0.0& 0.0& 700& -6.19& -7.57\\
LMC067982 &M2.5    I&3575&0.65& 0.0&-0.3&1040& -6.39& -8.27\\
LMC068098 &M1.5    I&3700&0.56& 0.0& 0.0& 650& -5.95& -7.40\\
LMC068125 &M4      I&3475&0.84& 0.0&-0.3&1080& -5.91& -8.21\\
LMC109106 &M2.5    I&3550&0.37& 0.0&-0.2& 890& -5.91& -7.89\\
LMC116895 &M0      I&3750&0.71& 0.0&-0.2& 880& -6.78& -8.10\\
LMC119219 &M3      I&3575&0.25&-0.5&-0.3&1150& -6.61& -8.48\\
LMC131735 &K2      I&4150&0.77& 0.0& 0.2& 510& -6.62& -7.36\\
LMC134383 &M2.5    I&3575&0.77& 0.0&-0.1& 800& -5.81& -7.69\\
LMC135720 &M4.5    I&3425&0.90&-0.5&-0.4&1200& -5.83& -8.38\\
LMC136042 &M3      I&3500&0.09&-0.5&-0.4&1240& -6.35& -8.54\\
LMC137624 &M0      I&3700&0.40& 0.0& 0.1& 600& -5.74& -7.19\\
LMC137818 &M2      I&3625&0.71& 0.0&-0.1& 740& -5.88& -7.57\\
LMC138405 &M1      I&3675&0.40& 0.0& 0.0& 650& -5.82& -7.35\\
LMC140296 &M2      I&3625&1.15& 0.0&-0.2& 990& -6.53& -8.22\\
LMC141430 &M1      I&3700&0.90&-0.5&-0.3&1110& -7.10& -8.55\\
LMC142202 &M1.5    I&3650&0.40&-0.5&-0.3&1050& -6.75& -8.36\\
LMC142907 &M2      I&3650&0.68& 0.0&-0.1& 790& -6.13& -7.74\\
LMC143877 &K3      I&3900&0.90& 0.0&-0.2&1010& -7.58& -8.58\\
LMC146126 &K5      I&3875&0.25& 0.0&-0.2&1050& -7.58& -8.62\\
LMC147199 &M1.5    I&3675&0.53& 0.0&-0.1& 810& -6.30& -7.82\\
LMC149721 &M0      I&3750&0.40& 0.0& 0.0& 670& -6.19& -7.51\\
LMC157533 &K5      I&3825&0.53& 0.0& 0.2& 510& -5.87& -7.00\\
LMC158317 &M1      I&3675&0.77& 0.0& 0.0& 680& -5.92& -7.45\\
LMC159974 &K1      I&4300&1.24& 0.0& 0.2& 560& -7.02& -7.72\\
LMC169754 &K2      I&4100&1.95& 0.0& 0.0& 700& -7.24& -8.01\\
LMC174714 &M1.5    I&3625&1.33& 0.0&-0.3&1080& -6.70& -8.39\\
LMC175464 &M2      I&3625&1.33&-0.5&-0.4&1200& -6.93& -8.62\\
LMC175746 &M3      I&3550&1.18& 0.0&-0.3&1100& -6.38& -8.35\\
LMC176890 &M0      I&3750&0.56& 0.0& 0.0& 670& -6.21& -7.52\\
LMC177150 &M1.5    I&3600&0.77& 0.0& 0.0& 650& -5.47& -7.26\\
LMC177997 &M1.5    I&3675&0.77& 0.0&-0.2& 980& -6.71& -8.24\\
\enddata
\end{deluxetable}

\begin{deluxetable}{l l c c c c c r c  r c c c r c r}
\rotate
\tabletypesize{\scriptsize}
\tablewidth{0pc}
\tablenum{3}
\tablecolumns{16}
\tablecaption{\label{tab:bb} Results from Broad Band Photometry}
\tablehead{
& Spectral
&\multicolumn{2}{c}{Spectral Fit\tablenotemark{a}}
&
&\multicolumn{5}{c}{Results from $(V-K)_0$}
&
&\multicolumn{5}{c}{Results from $(V-R)_0$}  \\ \cline{3-4} \cline{6-10} \cline{12-16} 
\colhead{Star}
&\colhead{Type}
&\colhead{$T_{\rm eff}$}
&\colhead{$M_{\rm bol}$}
&
&\colhead{$(V-K)_0$\tablenotemark{b}}
&\colhead{$T_{\rm eff}$}
&\colhead{$\Delta T_{\rm eff}$\tablenotemark{c}}
&\colhead{$M_{\rm bol}$}
&\colhead{$\Delta M_{\rm bol}$\tablenotemark{d}}
&
&\colhead{$(V-R)_0$\tablenotemark{e}}
&\colhead{$T_{\rm eff}$}
&\colhead{$\Delta T_{\rm eff}$\tablenotemark{c}}
&\colhead{$M_{\rm bol}$}
&\colhead{$\Delta M_{\rm bol}$ \tablenotemark{d}} 
}
\startdata
SMC005092&M2      I&3475& -8.48&&  4.42&3639&-163& -8.01& -0.47&&  1.10&3504& -28& -8.37& -0.11\\
SMC008930&M0      I&3625& -8.38&&  3.81&3878&-252& -7.94& -0.44&&  0.95&3726&-100& -8.15& -0.23\\
SMC010889&M0      I&3600& -8.47&&  4.29&3684& -83& -8.30& -0.17&&  1.04&3587&  13& -8.50&  0.03\\
SMC011101&K2.5    I&4200& -7.55&&  2.98&4306&-105& -7.43& -0.12&&  0.72&4215& -14& -7.54& -0.01\\
SMC011709&K5-M0   I&3725& -7.93&&  3.76&3898&-172& -7.69& -0.24&&  0.92&3779& -53& -7.82& -0.11\\
SMC011939&K2      I&4025& -8.05&&  3.23&4165&-139& -7.91& -0.14&&  0.80&4022&   3& -8.05&  0.00\\
SMC012322&K5-M0   I&3750& -8.34&&  3.61&3969&-218& -8.05& -0.29&&  0.92&3778& -27& -8.28& -0.06\\
SMC013740&K3      I&3750& -7.09&&  3.72&3916&-165& -6.87& -0.22&&  0.90&3830& -79& -6.94& -0.15\\
SMC013951&K3      I&4225& -7.76&&  2.89&4359&-133& -7.61& -0.15&&  0.72&4218&   7& -7.77&  0.01\\
SMC015510&K5      I&3850& -8.13&&  3.40&4072&-221& -7.88& -0.25&&  0.82&3979&-128& -7.96& -0.17\\
SMC018136&M0      I&3575& -8.76&&  3.99&3799&-223& -8.30& -0.46&&  0.99&3662& -86& -8.52& -0.24\\
SMC020133&M0      I&3625& -8.39&&  3.93&3826&-200& -8.04& -0.35&&  0.99&3669& -43& -8.28& -0.11\\
SMC021362&K5      I&3775& -7.53&&  3.79&3885&-109& -7.41& -0.12&&  0.90&3816& -40& -7.46& -0.07\\
SMC021381&K5      I&3800& -7.59&&  3.50&4020&-219& -7.33& -0.26&&  0.87&3885& -84& -7.46& -0.13\\
SMC023401&K1      I&4075& -7.18&&  3.15&4209&-133& -7.05& -0.13&&  0.76&4105& -29& -7.15& -0.03\\
SMC023743&K2      I&4050& -7.06&&  3.29&4130& -79& -6.99& -0.07&&  0.79&4040&  10& -7.07&  0.01\\
SMC025879&M0      I&3700& -8.44&&  3.37&4086&-385& -7.90& -0.54&&  0.87&3870&-169& -8.13& -0.31\\
SMC030135&K2      I&4050& -7.23&&  3.04&4269&-218& -7.02& -0.21&&  0.73&4194&-143& -7.11& -0.12\\
SMC030616&K2      I&3850& -8.22&&  3.47&4038&-187& -8.02& -0.20&&  0.84&3930& -79& -8.11& -0.11\\
SMC034158&K2      I&4075& -7.88&&  3.03&4278&-202& -7.68& -0.20&&  0.78&4070&   5& -7.88&  0.00\\
SMC035445&K1      I&4100& -7.66&&  3.13&4220&-119& -7.54& -0.12&&  0.79&4053&  47& -7.70&  0.04\\
SMC042438&K2      I&4250& -7.41&&  2.91&4350& -99& -7.29& -0.12&&  0.68&4308& -57& -7.37& -0.04\\
SMC043219&K2      I&3850& -7.26&&  3.64&3952&-101& -7.17& -0.09&&  0.89&3846&   4& -7.27&  0.01\\
SMC045378&K3      I&3850& -7.58&&  3.46&4044&-193& -7.37& -0.21&&  0.83&3958&-107& -7.43& -0.15\\
SMC046497&K5-M0   I&3700& -8.30&&  3.70&3924&-223& -7.96& -0.34&&  0.92&3785& -84& -8.12& -0.18\\
SMC046662&K3      I&4100& -8.08&&  3.40&4073&  27& -8.13&  0.05&&  0.83&3950& 150& -8.24&  0.16\\
SMC048122&K1      I&4225& -8.26&&  2.69&4491&-265& -7.99& -0.27&&  0.74&4171&  54& -8.30&  0.04\\
SMC049478&K5-M0   I&3700& -8.49&&  3.85&3858&-157& -8.26& -0.23&&  0.93&3775& -74& -8.34& -0.15\\
SMC050028&K1      I&4300& -9.14&&  2.51&4609&-308& -8.83& -0.31&&  0.71&4232&  68& -9.19&  0.05\\
SMC050840&M1      I&3625& -8.12&&  3.97&3807&-181& -7.80& -0.32&&  0.98&3676& -50& -8.00& -0.12\\
SMC054708&K1      I&4300& -7.76&&  2.81&4412&-111& -7.61& -0.15&&  0.72&4206&  94& -7.83&  0.07\\
SMC055681&K5-M0   I&4100& -8.40&&  2.83&4398&-297& -8.11& -0.29&&  0.74&4172& -71& -8.34& -0.06\\
SMC056732&K5-M0   I&3725& -7.66&&  3.72&3917&-191& -7.39& -0.27&&  0.89&3835&-109& -7.46& -0.20\\
SMC057386&K1      I&4300& -7.74&&  2.91&4346& -45& -7.66& -0.08&&  0.68&4301&   0& -7.74&  0.00\\
SMC057472&K2      I&4100& -7.60&&  3.20&4181& -80& -7.52& -0.08&&  0.76&4122& -21& -7.58& -0.02\\
SMC059803&K2-3    I&4100& -8.69&&  3.00&4294&-193& -8.50& -0.19&&  0.80&4016&  84& -8.78&  0.09\\
SMC060447&K2      I&3900& -7.37&&  3.41&4067&-166& -7.21& -0.16&&  0.84&3928& -27& -7.34& -0.03\\
SMC067509&K1      I&4175& -7.50&&  3.02&4284&-108& -7.38& -0.12&&  0.75&4129&  46& -7.54&  0.04\\
SMC069886&K5-M0   I&3750& -8.76&&  3.64&3952&-201& -8.49& -0.27&&  0.99&3672&  78& -8.93&  0.17\\
LMC054365&M2.5    I&3525& -7.88&&  4.39&3651&-125& -7.38& -0.50&&  0.99&3633&-107& -7.46& -0.42\\
LMC061753&M2      I&3600& -7.80&&  4.48&3623& -22& -7.68& -0.12&&  1.03&3581&  19& -7.87&  0.07\\
LMC062090&M1      I&3700& -7.92&&  3.92&3812&-111& -7.70& -0.22&&  0.91&3767& -66& -7.74& -0.18\\
LMC064048&M2.5    I&3500& -7.81&&  4.84&3532& -31& -7.57& -0.24&&  1.11&3483&  17& -7.88&  0.07\\
LMC065558&M1      I&3725& -7.57&&  3.91&3816& -90& -7.42& -0.15&&  0.95&3698&  27& -7.65&  0.08\\
LMC067982&M2.5    I&3575& -8.27&&  4.02&3774&-198& -7.70& -0.57&&  0.97&3674& -98& -7.92& -0.35\\
LMC068098&M1.5    I&3700& -7.40&&  4.09&3749& -48& -7.31& -0.09&&  0.93&3727& -26& -7.33& -0.07\\
LMC068125&M4      I&3475& -8.21&&  4.32&3671&-195& -7.44& -0.77&&  1.04&3569& -93& -7.81& -0.40\\
LMC109106&M2.5    I&3550& -7.89&&  4.00&3779&-228& -7.21& -0.68&&  0.95&3700&-149& -7.36& -0.53\\
LMC116895&M0      I&3750& -8.10&&  3.53&3981&-230& -7.75& -0.35&&  0.90&3795& -44& -7.98& -0.12\\
LMC119219&M3      I&3575& -8.48&&  3.88&3826&-250& -7.82& -0.66&&  0.93&3729&-153& -7.98& -0.50\\
LMC131735&K2      I&4150& -7.36&&  3.01&4256&-105& -7.28& -0.08&&  0.74&4126&  24& -7.37&  0.01\\
LMC134383&M2.5    I&3575& -7.69&&  4.73&3557&  18& -7.67& -0.02&&  1.06&3540&  35& -7.83&  0.14\\
LMC135720&M4.5    I&3425& -8.38&&  5.01&3494& -68& -7.92& -0.46&&  1.18&3421&   4& -8.40&  0.02\\
LMC136042&M3      I&3500& -8.54&&  4.83&3533& -32& -8.29& -0.25&&  1.07&3529& -28& -8.41& -0.13\\
LMC137624&M0      I&3700& -7.19&&  3.97&3792& -91& -7.01& -0.18&&  0.94&3710&  -9& -7.16& -0.03\\
LMC137818&M2      I&3625& -7.57&&  4.46&3631&  -5& -7.52& -0.05&&  1.07&3538&  87& -7.90&  0.33\\
LMC138405&M1      I&3675& -7.35&&  4.00&3781&-105& -7.11& -0.24&&  0.94&3710& -34& -7.24& -0.11\\
LMC140296&M2      I&3625& -8.22&&  3.90&3819&-193& -7.75& -0.47&&  0.96&3681& -55& -8.04& -0.18\\
LMC141430&M1      I&3700& -8.55&&  3.97&3792& -91& -8.37& -0.18&&  1.07&3533& 167& -9.14&  0.59\\
LMC142202&M1.5    I&3650& -8.36&&  4.19&3714& -63& -8.18& -0.18&&  0.95&3693& -42& -8.22& -0.14\\
LMC142907&M2      I&3650& -7.74&&  3.95&3799&-148& -7.39& -0.35&&  0.93&3732& -81& -7.49& -0.25\\
LMC143877&K3      I&3900& -8.58&&  3.00&4265&-364& -8.23& -0.35&&  0.78&4039&-138& -8.39& -0.19\\
LMC146126&K5      I&3875& -8.62&&  2.92&4313&-437& -8.19& -0.43&&  0.79&4008&-132& -8.43& -0.19\\
LMC147199&M1.5    I&3675& -7.82&&  4.75&3551& 124& -8.18&  0.36&&  1.10&3499& 176& -8.49&  0.67\\
LMC149721&M0      I&3750& -7.51&&  3.72&3893&-142& -7.28& -0.23&&  0.89&3797& -46& -7.39& -0.12\\
LMC157533&K5      I&3825& -7.00&&  3.81&3853& -27& -7.03&  0.03&&  0.89&3806&  19& -7.05&  0.05\\
LMC158317&M1      I&3675& -7.45&&  4.12&3736& -60& -7.30& -0.15&&  0.97&3662&  13& -7.49&  0.04\\
LMC159974&K1      I&4300& -7.72&&  2.68&4473&-172& -7.50& -0.22&&  0.67&4315& -14& -7.72&  0.00\\
LMC169754&K2      I&4100& -8.01&&  3.05&4232&-131& -7.92& -0.09&&  0.76&4086&  14& -8.01&  0.00\\
LMC174714&M1.5    I&3625& -8.39&&  4.05&3762&-136& -8.03& -0.36&&  0.96&3688& -62& -8.19& -0.20\\
LMC175464&M2      I&3625& -8.62&&  4.13&3734&-108& -8.32& -0.30&&  0.97&3672& -46& -8.47& -0.15\\
LMC175746&M3      I&3550& -8.35&&  4.43&3637& -86& -8.00& -0.35&&  1.04&3575& -24& -8.26& -0.09\\
LMC176890&M0      I&3750& -7.52&&  3.74&3885&-134& -7.32& -0.20&&  0.90&3779& -28& -7.45& -0.07\\
LMC177150&M1.5    I&3600& -7.26&&  4.38&3652& -51& -7.05& -0.21&&  1.05&3552&  48& -7.44&  0.18\\
LMC177997&M1.5    I&3675& -8.24&&  4.11&3740& -64& -8.08& -0.16&&  0.93&3727& -51& -8.09& -0.15\\
\enddata
\tablenotetext{a}{From Table~\ref{tab:results}.}
\tablenotetext{b}{$(V-K)_0=(V-K_S)-0.06-0.88A_V$.}
\tablenotetext{c}{$\Delta T_{\rm eff}$ = Effective temperature
adopted from spectral fitting the optical spectrophotometry {\it minus}
the effective temperature determined from the broad-band colors.}
\tablenotetext{d}{$\Delta M_{\rm bol}$ = Bolometric luminosity from
spectral fitting the optical spectrophotometry {\it minus} the bolometric
magnitude determined from the broad-band colors.}
\tablenotetext{e}{$(V-R)_0=(V-R)-0.19A_V$.}
\end{deluxetable}

\clearpage
\begin{deluxetable}{l c r r c c c r r c c c r r c }
\pagestyle{empty}
\tabletypesize{\footnotesize}
\tablewidth{0pc}
\tablenum{4}
\tablecolumns{15}
\tablecaption{\label{tab:tscale}Effective Temperature Scales}
\tablehead{
\colhead{} &
\multicolumn{4}{c}{SMC} & &
\multicolumn{4}{c}{LMC} &&
\multicolumn{4}{c}{Milky Way\tablenotemark{a}} \\ \cline{2-5} \cline{7-10} \cline{12-15}
\colhead{Spectral} & 
\colhead{$T_{\rm eff}$}  & 
\multicolumn{4}{c}{} & 
\colhead{$T_{\rm eff}$}  & 
\multicolumn{4}{c}{} & 
\colhead{$T_{\rm eff}$} & \\
\colhead{Type} &
\colhead{(K)} &
\colhead{$\sigma_\mu$\tablenotemark{b}} &
\colhead{$N$} &
\colhead{BC} & &
\colhead{(K)} &
\colhead{$\sigma_\mu$\tablenotemark{b}} &
\colhead{$N$} &
\colhead{BC} & &
\colhead{(K)} &
\colhead{$\sigma_\mu$\tablenotemark{b}} &
\colhead{$N$} &
\colhead{BC} \\
}
\startdata
K1-K1.5 I & 4211 & 34 & 7  & -0.73 && 4300 & \nodata & 1 & -0.70 &&  4100 & 100 & 3 & -0.79\\
K2-K3 I    & 4025 & 38 & 15& -0.92 && 4050 & 62          & 3 & -0.80 && 4015 & 40 & 7 & -0.90 \\
K5-M0 I   & 3788 & 36 & 10 & -1.27 && 3850 & 18         & 2 & -1.09 && 3840 & 30 & 3 &-1.16 \\
M0 I         & 3625 & 19 & 5 &   -1.62  && 3738 & 11         & 4 & -1.31 && 3790 & 13 & 4 & -1.25\\
M1 I         & 3625 & \nodata & 1 &-1.61 && 3695 & 8 & 5  &  -1.45  && 3745 & 17 & 7 & -1.35 \\
M1.5 I      & \nodata & \nodata & \nodata &\nodata && 3654 & 14 & 6 & -1.59 && 3710  & 8 & 6 & -1.43 \\
M2 I         & 3475 & \nodata & 1 & -2.07  && 3625 & 7 & 5 & -1.69 && 3660 & 7 & 17 & -1.57 \\
M2.5 I     & \nodata & \nodata & \nodata & \nodata && 3545 & 13 & 5 & -1.99 && 3615 & 10 & 5 & -1.70 \\
M3  I        & \nodata & \nodata & \nodata & \nodata && 3542 & 18 & 3 & -2.01 && 3605  & 4 & 9 & -1.74 \\
M3.5 I     & \nodata & \nodata & \nodata & \nodata && \nodata & \nodata & \nodata & \nodata &&3550 & 11 & 6 & -1.96 \\
M4-M4.5 I    & \nodata & \nodata & \nodata & \nodata && 3450 & \nodata & 2 & -2.18 && 3535 & 8 & 6 & -2.03 \\
M5 I & \nodata & \nodata & \nodata & \nodata && \nodata & \nodata & \nodata & \nodata && 3450 & \nodata & 1 & -2.49 \\
\enddata
\tablenotetext{a}{From Paper I.}
\tablenotetext{b}{Standard deviation of the mean.}
\end{deluxetable}

\clearpage

\end{document}